\documentclass[fleqn,usenatbib]{mnras}

\usepackage[T1]{fontenc}
\usepackage{ae,aecompl}
\def\be{\begin{equation}}
\def\ee{\end{equation}}
\def\ba{\begin{eqnarray}}
\def\ea{\end{eqnarray}}

\def\init{_I} 
\def\now{_F}  
\def\gpot{\phi}          
\def\dgpot{\psi}         
\def\cpot{\Phi}          
\def\dcpot{\Psi}         
\def\bx{{\bf x}}         
\def\bq{{\bf q}}         
\def\bv{{\bf v}}         
\def\bp{{\bf p}}         
\def\bg{{\bf g}}         
\def\everything{V}       

\usepackage[utf8]{inputenc} 
\usepackage[T1]{fontenc}

\usepackage{graphicx}	

\usepackage{amsmath}	
\usepackage{amssymb}	

\title[Semi-discrete optimal transport reconstruction algorithm]{A fast semi-discrete optimal transport algorithm for a unique reconstruction of the early Universe}

\author[B.~{Levy}, R.~{Mohayaee}, S.~{von Hausegger}]{
Bruno Levy$^{1}$\thanks{E-mail: bruno.levy@inria.fr},
Roya Mohayaee$^{2}$\thanks{E-mail: mohayaee@iap.fr }
and Sebastian von Hausegger$^{1,2,3}$\thanks{E-mail: sebastian.vonhausegger@physics.ox.ac.uk}
\\
$^{1}$Universit\'es de Lorraine, CNRS, Inria, LORIA, F-54000 Nancy, France\\
$^{2}$Sorbonne Universit\'e, CNRS, Institut d'Astrophysique de Paris, 98bis Bld Arago, 75014 Paris, France\\
$^{3}$Rudolf Peierls Centre for Theoretical Physics, University of Oxford, Parks Road, Oxford, OX1 3PU, United Kingdom
}

\date{Accepted XXX. Received YYY; in original form ZZZ}

\pubyear{2020}

\begin{document}
\label{firstpage}
\pagerange{\pageref{firstpage}--\pageref{lastpage}}
\maketitle
\begin{abstract}
We leverage powerful mathematical tools stemming from optimal transport theory and transform them into an efficient algorithm to reconstruct the fluctuations of the primordial density field, built on solving the Monge-Amp\`ere-Kantorovich equation.
Our algorithm computes the optimal transport between an initial uniform continuous density field, partitioned into Laguerre cells, and a final input set of discrete point masses, linking the early to the late Universe. While existing early universe reconstruction algorithms based on fully discrete combinatorial methods are limited to a few hundred thousand points, our algorithm scales up well beyond this limit, since it takes the form of a well-posed smooth convex optimization problem,
solved using a Newton method. We run our algorithm on cosmological $N$-body simulations, from the AbacusCosmos suite, and reconstruct the initial positions of $\mathcal{O}(10^7)$ particles within a few hours with an off-the-shelf personal computer. We show that our method allows a unique, fast and precise recovery of subtle features of the initial power spectrum, such as the baryonic acoustic oscillations.
\end{abstract}



\section{Introduction}
    Optimal Transport theory has found spectacular applications in diverse areas of science, from economics to biology, physics, data science and machine learning to name but a few. The emerging applications of the two-centuries-old theory 
    is mainly due to advances in mathematics and the developments of fast new algorithms.
    It is indeed these fundamental advances that have paved the way for major breakthroughs in artificial intelligence, since they made it possible to compute a natural, \emph{Wasserstein}, distance between entities of various nature, essential for object recognition and classification.
    
   The reason behind the success of optimal transport theory in physics might be that it describes a universal foundation of Nature, where most processes seem to be governed by the optimisation of an underlying, sometimes unknown, quantity.  Light follows a path that, roughly, minimises time by Fermat's principle, and freely moving test particles follow time-like geodesics in general relativity.
    As perhaps best phrased by Euler more than two centuries ago: “nothing arises in the universe in which one cannot see the sense of some maximum or minimum." The variational principle founded by Euler himself and at the basis of the classical mechanics and quantum field theory has now become a subset of the vast field of variational calculus. 
    
    The Euler-Lagrange action optimisation, often referred to as the least action principle, found its application in cosmology, thanks to the pioneering work of Peebles back in 1989.   
     Peebles aimed at reconstructing the past history of the Local Group by retracing the trajectories of around 10 member-galaxies back in time \citep{1989ApJ...344L..53P}. He showed that the recovery of the initial conditions without prior knowledge of present velocities is possible by considering that, at early times, the peculiar velocities of matter is negligible and their spatial distribution uniform. He thus solved a mixed boundary-value problem instead of the usual Cauchy problem \citep{1989ApJ...344L..53P}.
    The method not only provided valuable informations on the orbits of the members of the Local Group but also put constraints on cosmological parameters. Indeed a low-value of the matter density parameter was favoured by Peebles' action minimisation at a time when the Cold Dark Matter (CDM) model was the standard paradigm. Peebles' algorithm was made more efficient for application to larger datasets but lacked uniqueness: existence of mulitple minima, maxima and saddle points in the landscape of solutions lead to multiple trajectories, all physically viable ~\citep{1994ApJ...429...43P,1995ApJ...454...15S,2000MNRAS.313..587N,2002MNRAS.335...53B}. Most recently, the method's excess complexity has been reduced in new numerical schemes and also by smoothing out strongly non-linear scales \citep{2019MNRAS.484.3818S,2020arXiv201010456S}.

Most action optimisations involve finding minimum energy trajectories 
for fixed end points. The added complexity of the cosmological setting in Peebles' formulation is that in addition to the trajectories, the initial positions of galaxies are also unknown, which renders the problem highly under-determined. Can we add suitable constraints on the trajectories to achieve a unique solution ?
On large scales\footnote{Unlike fluid mechanics, cosmology lacks a proper control parameter, e.g. a Reynolds number, and we can only separates single from multistream regimes in an empirical manner.}-- roughly scales of tens of megaparsecs--the velocities of galaxies, as tracers of the underlying dark matter fluid, remain a potential flow, as has been shown numerically by N-body simulations and theoretically at least up to the third-order in the Lagrangian perturbation theory ({\it e.g.} see \citet{1995MNRAS.276..115C}). 
 In previous works, we showed that where the trajectories of the fluid elements have not crossed or when their velocities is the gradient of a convex potential, the cosmological reconstruction problem is a well-posed instance of what is called \emph{the optimal mass transportation problem} and has a unique solution \citep{EURNature,EUR}. \\

The mass transportation problem\footnote{which historically preceds the variational calculus of Euler and Lagrange mentioned at the beginning of this introduction.}, was initially formulated by Monge during the French revolution. Monge aimed at finding how to transport soil from N number of excavated holes to the same number of rubbles while minimizing the total product of the transported mass and the travelled distance \citep{Monge1784}. This problem has a rich mathematical
structure that was revealed later by a continuous stream of advances both in fundamental mathematics (see review by \cite{OTON}) and in applied mathematics (see review by \cite{ComputationalOT}). The most prominent "quantum leap" was made during WWII by Kantorovich, who invented the "mathematical toolbox" to study the existence and uniqueness \citep{Kanto:1942}. Kantorovich studied a relaxation of the problem\footnote{where the unknown is the "graph" of the function in the product space, referred to as the "transport plan". More on this in \S \ref{sec:dualMAK}.}, and introduced the dual formulation, with Lagrange multipliers. The relaxed problem was subsequently referred to as the Monge-Kantorovich mass transportation problem. From our cosmological perspective, these Lagrange multipliers correspond to the initial and final gravitational potentials, related to each other by the Legendre-Fenchel transform. \\
   
A more recent "quantum leap" was made by Brenier with his celebrated \emph{polar factorization theorem}, that states that the optimal-transport map corresponds to the gradient of a convex potential \citep{BrenierPFMR91}. By injecting the gradient of the potential into the mass conservation constraint of the Monge-Kantorovich problem, one obtains a non-linear partial-differential equation (PDE) known as the \emph{Monge-Ampère equation}, that can be solved to find the potential\footnote{Interestingly, this very class of PDE was also first studied by Monge during the French revolution and later generalised by Ampère at the beginning of the 1800's.}. After the polar factorization theorem was discovered, \cite{DBLP:journals/nm/BenamouB00} revealed that the minimized quantity, the \emph{Wasserstein distance}, corresponds to the \emph{action integral} in an incompressible Euler fluid. 
    
In our cosmological setting, it also corresponds to the action integral in a simple model of self-gravitating matter--that is the integrated kinetic energy. These theorems give us a way of computing the potential, and reconstructing the trajectories. For example, in the Zel'dovich regime, each
fluid element follows a rectilinear motion with a constant speed that corresponds to the
gradient of the reconstructed primordial gravitational potential. In the language of optimal transport, the gravitational potential can be deduced from the Lagrange multiplier of the mass conservation constraint, through the Legendre-Fenchel transform. This characterization implies that the potential is a convex function. In terms of physics, it implies that there is no multistreaming in the reconstructed dynamics.  \\

The Monge-Ampère-Kantorovich (MAK) cosmological reconstruction method was developed based on these advances in optimal transport theory and subsequently solved using a fully-discrete combinatoric algorithm \citep{EURNature,EUR}.
The algorithm was tested on simulations  \citep{2003A&A...406..393M, 2006MNRAS.365..939M,2008PhyD..237.2145M,2008MNRAS.383.1292L} and also applied to galaxy redshift surveys \citep{2005ApJ...635L.113M,2010ApJ...709..483L} and found applications in condensed-matter physics ({\it e.g.} \citet{2012JSP...147..487A}).
However, its cubic algorithmic complexity rendered it impractical for applications 
to challenging and forefront cosmological problems such as the reconstruction of the sound horizon at decoupling, {\it i.e.} the scale of the baryon acoustic oscillations (BAO) \citep{2005ApJ...633..560E,2007ApJ...664..675E}. 

In this article, we design a highly-efficient new algorithm, which yields a \emph{unique} solution by construction and can be applied to computationally-demanding problems in cosmology. We demonstrate the efficiency and accuracy of our algorithm through the example of BAO reconstruction. In our case, the initial condition is a uniform density field (continuous), and the final one corresponds to a set of galaxies, represented by a (discrete) set of points, hence a {\it semi-discrete} optimal transport problem. The theory of optimal transport is written in a mathematical language (theory of probability measures) that is general enough to encompass such irregular settings (with mass concentrated on points). Not only this mathematical language is exactly what we need to model our cosmological problem, but also it can be directly translated into a computational algorithm that can be efficiently implemented on a computer: 
solving the underlying semi-discrete Monge-Ampère equation is equivalent
to minimizing a \emph{smooth} and \emph{convex} function that depends on the final gravitational potential at each discrete point. It is much faster than
solving a \emph{discrete} Monge-Ampère equation, as we did before, that required exploring a huge combinatorial space \citep{EURNature,EUR}.
The convergence between the three aspects of the problem (cosmology, mathematics, computer science) results in a new algorithm that solves the assignment problem with an empirical complexity of $N{\rm log}N$ (as compared to $N^3$), and that can be applied to reconstruction problems of unprecedented sizes: $O(10^6)$ particles in \emph{minutes}, to be compared with
\emph{months}, and $O(10^8)$ particles in hours. Using our semi-discrete algorithm on a set of N-body simulations taken from Abacus Cosmos suite \citep{Garrison2018,Garrison2019}, we examine the complexity of our algorithm and compare reconstructed and simulated initial density fields, their power spectra and correlation functions, starting from two different redshifts. We show that these quantities can be accurately reconstructed above scales of a few Mpcs with competitive numerical speed. In particular, we show that BAO can be reconstructed with high accuracy and speed both in the power spectrum and the correlation function. \\

Indeed, the BAO scale has much decisional power, by providing a rather robust standard ruler of cosmology. As a powerful distance indicator, BAO measurement can probe the acceleration phase in the expansion history of the Universe and distinguish between the theory of general relativity and those of modified gravity.
It is also a pleasantly obvious feature of the two-point correlation function of the galaxy field. Non-linear gravitational evolution at late times slightly softens this feature and therewith the statistical certainty with which the BAO scale can be determined. Beyond general (but not less important) questions about, {\it e.g.}, statistical properties of the initial density fluctuations, un-doing the non-linear evolution to obtain the linear density field from low-redshift measurements of the large-scale structure give strong motivation for reconstruction techniques ({\it e.g.} \citet{2005ApJ...633..560E,2007ApJ...664..675E,2010ApJ...720.1650S,2012MNRAS.427.2132P}). 
Since the pioneering variational method of Peebles, numerous reconstruction techniques, for different tasks and not just the BAO retrieval, have been proposed. Here we can only mention a few. Many of these methods take a probabilistic approach
({\it e.g.} \cite{1992MNRAS.254..315W,kitaura2008,2009PhRvD..80j5005E,2011ApJ...731..116N,2014MNRAS.441.2923C,2013MNRAS.432..894J,2019MNRAS.488.2573B}), a few others are perturbative  ({\it e.g.} \cite{1992ApJ...391..443N,1993ApJ...405..449G,1998ApJ...492....1K,2007ApJ...664..675E,2017PhRvD..96b3505S}) and many are variational
({\it e.g.} \cite{1997MNRAS.285..793C,1999ApJ...515..471N,1999MNRAS.308..763M,2017ApJ...841L..29W,2018PhRvD..97b3505S}). We refer the reader to Section 7 of \cite{EUR} for a detailed discussion of these categories of reconstruction methods. Here we present our algorithm that, by contrast to many of the other methods, yields a unique solution by construction, is deterministic, model-independent and efficient computationally.
\\

This article is structured as follows: we first provide a review of the Lagrangian dynamics in a expanding Universe and show the generality and limitations of our two hypothesis of gradient flow and convexity (Section \ref{sec:MAK}). 
Then we explain old methods and our new algorithm to solve the underlying assignment problem (Section \ref{sec:assign}), before giving the details of our numerical solution mechanism (Section \ref{sec:num}). Finally we test the algorithm against the \textsc{AbacusCosmos} simulations in Section \ref{sec:results}. In section \ref{sec:conclusion} we conclude.

\section{Monge-Amp\`ere-Kantorovich (MAK) reconstruction}
\label{sec:MAK}

\subsection{Problem setting}

We consider the problem of reconstructing the fluctuations in the initial condition of self-gravitating matter governed by the following equations, in Eulerian form (see \cite{Peebles:1980}, or Appendix A of \cite{EUR}):
\begin{align}
    \label{eq:momentum} 
        \partial_\tau \bv + (\bv\cdot\nabla_{x})\bv &= -\frac{3}{2\tau} (\nabla_{x} \gpot + \bv ) \\
    \label{eq:continuity}
         \partial_\tau \rho + \nabla_{x}\cdot(\rho \bv)     &= 0 \\
    \label{eq:poisson}
          \nabla^2_{xx} \gpot &= \frac{\rho - 1}{\tau}
\end{align}
were $\rho$ is the density field defined over $V \subset \mathbb{R}^3$, $\gpot$ the gravitational potential, $\tau$ is the growth rate of structures, used as a time variable, and normalized in such a way that $\tau={\tau\init}=0$ corresponds to the initial condition, and $\tau = \tau\now = 1$ corresponds to present time, and $\bx$ denotes the co-moving coordinates.
The peculiar velocity field $\bv$ is also expressed as a function of the co-moving coordinates $\bx$. 
 
Equation (\ref{eq:momentum}) is the momentum (Euler) equation. Its right-hand side has two terms that have opposite effects: the first term ($-\frac{3}{2\tau} \nabla_x\gpot$) corresponds to the effect of gravity, that tends to collapse structures, and the second term ($-\frac{3}{2\tau}\bv$), called the Hubble-Lema\^{\i}tre drag, corresponds to the effect of expansion, that slows down the collapsing effect of gravity. Equation (\ref{eq:continuity}) enforces mass conservation (continuity equation) and Equation (\ref{eq:poisson}) is the Poisson equation that governs the gravitational potential.

Given the density field  $\rho\now$ at time $\tau\now = 1$ that corresponds to the present distribution of mass, 
our goal is to reconstruct the initial fluctuations of  $\rho(.,\tau\init + \epsilon)$ for a small $\epsilon$.
One can also consider the problem of reconstructing the full dynamics of the system $\rho(\tau,\bx)$ and $\bv(\tau,\bx)$ for 
$\tau \in [\tau\init, \tau\now]$ and $\bx \in \everything$. Clearly, the problem is under-determined but, as we shall show later, under some reasonable simplifying assumptions, it can be replaced by a well-posed convex optimization problem. 
 
\subsection{Lagrangian perturbation theory}
\label{sec:HOPT}
One can observe that at the initial condition $\tau\init=0$, for the right hand side of the Poisson equation (\ref{eq:poisson}) to be defined, density needs to be uniform $\rho\init(.) = \rho(.,\tau\init)= 1$. The same consideration for the right-hand side of the momentum equation (\ref{eq:momentum}) implies that at the initial condition, the velocity coincides with the (negated) gradient of the potential $\bv\init(.) = \bv(.,\tau\init) = -\nabla_x\gpot(.,\tau\init)$. This condition, which also equivalently arises as a solution to the linearised set of equations (\ref{eq:momentum})-(\ref{eq:poisson}), is sometimes referred to as \emph{slaving}\footnote{Here the initial condition is considered from a mathematical point of view. From a physical point of view, clearly, there cannot be a non-uniform potential associated with a uniform density. In fact, it is at $\tau\init + \epsilon$ that the potential is non-uniform, but one can make this $\epsilon$ arbitrarily small. In a certain sense, one can "push" the non-uniform potential from $\tau\init + \epsilon$ towards $\tau\init$: the right-hand side of the Poisson equation for the potential (eq. \ref{eq:poisson}) with $\tau$ in the denominator results in a significant potential yielded by tiny fluctuations of the density at the initial condition.}
Consider now the Lagrangian point of view, and focus on the mass element that is at position $\bq$ at time
$\tau\init$. Denote its trajectory by $\bx(\bq,\tau)$. Its initial speed at time $\tau\init$ is given by 
$\bv\init(\bq) = -\nabla_q \gpot(\bq,\tau\init) = -\nabla_q \gpot\init(\bq)$. In 1D, one can prove that the speed of the mass
element remains constant at any time $\tau$ (see e.g. \cite{EUR} for the proof). In other words, integrating (\ref{eq:momentum}), ({\ref{eq:continuity}), (\ref{eq:poisson}) in 1D results in a uniform rectilinear motion for all mass elements:
\begin{align}
{\cal D}_\tau {\bf x}({\bf q},\tau) &= (1/\tau\now)({\bf x}\now({\bf q}) - {\bf q}) 
                              = -\nabla_{q}\ \gpot\init({\bf q}) \quad \forall \tau 
\label{eq:1D1}  \\                            
{\bf x}({\bf q},\tau) &= (1-\tau/\tau\now) {\bf q} + (\tau/\tau\now){\bf x}\now({\bf q})
\label{eq:1D}
\end{align}
where $\bx\now(\bq) = \bx(\bq,\tau\now)$  and where ${\cal D}_\tau$ denotes the Lagrangian derivative w.r.t. time $\tau$. It also means that in 1D, to determine the entire motion, one only needs to know the initial potential $\gpot\init(.)=\gpot(.,\tau\init)$ at time $\tau\init=0$.

In 3D, for a small time $\tau$, the speed of a mass element is still given by (\ref{eq:1D1}), but it is no longer strictly the case at any time. However it is considered to be a reasonable approximation \citep{Zeldo:1970}.  This approximation means that the r.h.s. of the momentum equation (\ref{eq:momentum}) vanishes. Physically, it means that the Hubble-Lema\^{\i}tre drag exactly counter-balances the effect of gravity, implying that each mass element has a uniform rectilinear motion (\ref{eq:1D}). In this setting, to reconstruct the full dynamics, one just needs to determine the $\everything \rightarrow \everything$ map $\bq \mapsto \bx\now(\bq)$. This map is in turn completely determined by the potential $\gpot\init(.) = \gpot(.,0)$, using the relation $\bx\now(\bq) = \bq + \tau\now \bv\init(\bq) = \bq - \tau\now \nabla_{q}\ \gpot\init(\bq)$.

At this point, given the potential $\gpot\init$ at the initial condition (we will see how to compute it in Section \ref{sec:num}), we can reconstruct the Zel'dovich approximation. This gives us for the mass particle that was located at $\bq$ at time $\tau\init$ its position $\bx\now(\bq) = \bq - \tau\now \nabla_q \ \gpot\init(\bq)$ at the present time $\tau\now = 1$. In other words, this gives us the \emph{assignment} between the initial condition and the present distribution of mass from which we can obtain the particle positions at arbitrary times ($\tau$), up to the first-order Lagrangian perturbation theory, as
\be
    \mathbf{x}(\tau,\mathbf{q}) =  \mathbf{q} + \frac{\tau}{\tau_F}(\bx\now(\bq)-\mathbf{q})\,.
   \label{eq:secondorder}
\ee

\noindent
Although the assignement between $\bq$ and $\bx\now$ is valid for as long as the convexity holds, we limit ourselves to the first-order only for obtaining the particle positions at 
intermediate times. The main reason is that here we test our method with the goodness of reconstruction of BAO.
It happens that often one adds
additional, broad band, fitting terms to the power spectrum which takes care of the mode-coupling as well as other effects such as the shot noise. The implementation of the second and higher-order Lagrangian perturbation theory into our algorithm shall be reported in the forthcoming works. 

\subsection{Least action principle and optimal transport}

One can also obtain the momentum equation (\ref{eq:momentum}) by extremizing the following action integral (\cite{EUR} Appendix D):
\begin{equation}
\label{eq:action1}
I = \frac{1}{2} \int_{\tau\init}^{\tau\now}\int_\everything \quad (\rho|\bv|^2 + \frac{3}{2}|\nabla_x \gpot |^2) \tau^{3/2} \quad d^3 \bx\ d\tau
\end{equation}
subject to mass conservation (\ref{eq:continuity}), to the Poisson equation for the potential (\ref{eq:poisson}) and to the boundary conditions:
\be
  \label{eq:action_bc}
  \rho(.,\tau\init) = \rho\init(.) = 1 \quad ; \quad \rho(.,\tau\now) = \rho\now(.)
\ee
where $\rho\init$ denotes the (uniform) density map at the initial condition and $\rho\now$ denotes the density map at present time $\tau = \tau\now$. Using the method of Lagrange multipliers and varying $\rho$, one obtains the momentum equation (\ref{eq:momentum}).  

Consider now an approximation, where the second term of the integrand and the $\tau^{3/2}$ factor are removed \citep{ZeldoAction}, which may be thought of as replacing the $3/2$ coefficient by $3 \alpha/2$ and making $\alpha$ tend to zero. The action integral (\ref{eq:action1}) then becomes:
\begin{equation}
\label{eq:action2}
I = \frac{1}{2} \int_{\tau\init}^{\tau\now}\int_\everything \quad \rho|\bv|^2 \quad d^3\bx\ d\tau.
\end{equation}

Note that the integrand has only the kinetic energy, and no longer any potential energy. This again corresponds to the Zel'dovich approximation.
Given the boundary conditions (\ref{eq:action_bc}), now we want to find the motion that minimizes the kinetic energy. If we have a single mass particle, it is easy to see that minimizing the action results in a uniform rectilinear motion \citep{LaudauCourse}. It can be proved \citep{DBLP:journals/nm/BenamouB00} that this is still the case for any number of particles, or even for a continuous density field $\rho$: extremizing the action (\ref{eq:action2}) means that all mass particles (or all elementary mass elements for a continuous $\rho$) move in a uniform rectilinear manner. In other words, finding the motion $\bx(\bq,\tau)$ that minimizes the action $I$ on $\everything \times [\tau\init,\tau\now]$ is equivalent to finding the map $\bx\now : \everything \rightarrow \everything$ that gives the position at present time $\bx\now(\bq) = \bx(\bq,\tau\now)$ of the mass element
that was at position $\bq$ at the initial condition $\tau\init=0$. The map $\bx\now$ minimizes the following functional:
\begin{equation}
    \inf_{\bx\now} \int_\everything \rho(\bq) | {\bf q} - \bx\now(\bq)|^2 \ d^3\bq
    \label{eq:OTcost}
\end{equation}
subject to mass conservation (\ref{eq:continuity}) and to the boundary conditions (\ref{eq:action_bc}). Now, it may be more natural to write mass conservation in Lagrangian coordinates. Using $\rho(\bx(\bq,\tau),\tau) = \rho(\bq)/(\mbox{det} \nabla_q \bx)$, the mass conservation constraint writes:
\begin{equation}
    \rho\now\left(\bx\now(\bq)\right)\ \mbox{det}\left(\nabla_q\ \bx\now(\bq)\right) = \rho\init(\bq) \quad \forall \bq
    \label{eq:OTcnstr}
\end{equation}

The minimization of expression (\ref{eq:OTcost}) subject to (\ref{eq:OTcnstr}) is referred to as \emph{Monge's optimal transport} problem \citep{Monge1784}.

\subsection{The convex Kantorovich potential and the Monge-Amp\`ere equation}

\label{sec:Kpot}

Consider Monge's optimal transport problem (\ref{eq:OTcost}). Introducing the Lagrange multiplier $\dcpot: \everything \mapsto \everything$ associated with the constraint (\ref{eq:OTcnstr}), and using the identity $|\bx\now(\bq) - \bq|^2 = \bx\now(\bq)^2 - 2 \bq \cdot \bx\now(\bq) + \bx^2$, the optimal transport problem can be written as the following saddle point problem:
\ba
    \nonumber
    \sup_{\bx\now} \inf_{\dcpot} \left[\  J(\bx\now)  = \int_\everything \rho\init(\bq) \bx\now(\bq) \cdot \bq \ d^3\bq \right. \\ 
       \left. -  \int_\everything \dcpot(\bx\now(\bq) \rho(\bq)\ d^3\bq  +   \int_\everything \dcpot(\bx) \rho\now(\bx)\ d^3\bx \ \right].
\ea \\
 
Formally, the first-order optimality condition w.r.t. $\bx\now$ writes:
\ba
  \nonumber
   \frac{\partial J}{\partial \bx\now} = 0 & \Rightarrow & \rho\init(\bq)\ \bq = \rho\init(\bq)\nabla\dcpot(\bx\now(\bq))  \\
   & \Rightarrow & \bq = \nabla_x \dcpot(\bx\now(\bq))
   \label{eq:conv1}
\ea

The second-order optimality condition writes:
\ba 
  \nonumber
   \frac{\partial^2 J}{\partial \bx\now^2} \ge 0  & \Rightarrow & D^2\dcpot \ge 0 \\
                                                   & \Rightarrow & \dcpot\ \mbox{is a convex function}
   \label{eq:conv2}
\ea
where $D^2\dcpot$ denotes the determinant of the Hessian matrix of $\dcpot$ (the derivations above (\ref{eq:conv1}), (\ref{eq:conv2}) just give an intuition, the reader is referred to \citep{BrenierPFMR91} for a rigorous proof).

From the first-order and second-order optimality conditions, we learn that the function $\bx \mapsto \bq$ that maps a mass element $\bx$ at present time $\tau\now$ back to its initial position $\bq$ at time $\tau\init$ is the gradient of a convex potential $\dcpot$ (called the \emph{Kantorovich potential}).
 
Next, we study the forward map $\bq \mapsto \bx\now$. It can be observed that the variables $\bq$ and $\bx_f$ play a symmetric role in the optimal transport problem. By exchanging the roles of $\bx\now \leftrightarrow \bq$, one can find a similar relation for the forward map $\bq \mapsto \bx\now(\bq)$:
\be
   \bx\now(\bq) = \nabla_q \cpot(\bq) \quad ; \quad \cpot\ \mbox{is a convex function}.
\ee
 
 \noindent
 It can be shown that this symmetry implies a relation between $\dcpot$ and $\cpot$: they are the Legendre-Fenchel transform\footnote{
The Legendre-Fenchel transform plays an important role in mechanics and thermodynamics, since it corresponds to the relation between Hamilton and Lagrange equations, see for instance \citep{LaudauCourse}, chapter 7.
} of each-other, given by:
\be
    \forall \bq, \cpot(\bq) = \dcpot ^*(\bq) \quad \mbox{where}\, \dcpot ^*(\bq) = \sup_{\bx} \left[\bx \cdot \bq - \dcpot(\bx) \right].
\ee
 
\noindent 
Next, we recall that the map $\bx\now(.)$ is determined by the gravitational potential $\gpot\init(.)$ at the initial condition by $\bx\now(\bq) = \bq - \nabla_q \gpot\init(\bq)$. This
gives us the relation between the gravitational potential $\gpot\init$ at time $\tau\init$ and the convex Kantorovich potential $\cpot$ associated with the map $\bq \mapsto \bx\now(\bq)$:
\ba
\nonumber
\bx\now(\bq) & = & \bq - \nabla_q \gpot\init(\bq) \\
\nonumber
              & = & \nabla_q \cpot(\bq) \\
              & \Rightarrow & \cpot(\bq) = 1/2\bq^2 - \gpot\init(\bq)
\ea

\noindent
The insertion of $\bx\now(\bq) = \nabla_q \phi(\bq)$ into the (Lagrangian) mass conservation constraint (\ref{eq:OTcnstr}) yields
\be
   \label{eq:MA}
   \rho\now\left(\nabla_x \cpot(\bq)\right) D^2\cpot(\bq) = \rho\init(\bq),
\ee
where $D^2\cpot(.)$ denotes the determinant of the Hessian matrix of $\cpot$. Equation (\ref{eq:MA}) is referred to as the \emph{Monge-Amp\`ere equation} (or MA equation for short). \\

In our context, the convexity of the Kantorovich potential has an interesting consequence: it implies that there is no multistreaming in the reconstructed dynamics.
It can be proven by contraction:
Consider two distinct points $\bq_1,\bq_2$ and their images $\bx\now(\bq_1), \bx\now(\bq_2)$ through the $\bx\now(.)$ map. They move along the following trajectories:
\ba
\nonumber
\bx_1(\tau) & = & (1-\tau) \bq_1 + \tau \bx\now(\bq_1) \\
\bx_2(\tau) & = & (1-\tau) \bq_2 + \tau \bx\now(\bq_2).  
\ea
If there was multistreaming, then there would exist a time $\tau$ such that $\bx_1(\tau)  =  \bx_2(\tau)$, or:
\ba
\nonumber
(1-\tau) \bq_1 + \tau \bx\now(\bq_1)  & = &  (1-\tau) \bq_2 + \tau \bx\now(\bq_2) \\ 
\nonumber
(1-\tau) \bq_1 + \tau \nabla_q \cpot(\bq_1) & = & (1-\tau) \bq_2 + \tau \nabla_q \cpot \varphi(\bq_2) 
\ea
\vspace{-7mm}
\ba
\nonumber
(1-\tau)(\bq_2 - \bq_1) + \tau\left(\nabla_q \cpot(\bq_2) - \nabla_q \cpot(\bq_1)\right) & = & 0 \\
\label{eq:convexity}
(1-\tau)(\bq_2 - \bq_1) + \tau\left(\nabla_q \cpot(\bq_2 - \bq_1)\right) & = & 0. 
\ea

\noindent
The last line (\ref{eq:convexity}) contradicts the convexity of $\cpot$, that implies that $(1-\lambda)(\bq_2 - \bq_1) +  \lambda\nabla_q\cpot(\bq_2 - \bq_1)$ is strictly greater than zero for all
$\bq_1 \neq \bq_2 \in \everything$ and $\lambda \in [0,1]\ \blacksquare$

\subsection{An overall account of this section}

To summarize, given the density at present time $\rho\now$ and the density at the initial condition $\rho\init = 1$, our goal is to find the 
assignment map $\bq \mapsto \bx\now(\bq)$ that determines the assignment between the points $\bq$ at the initial condition and the points $\bx$ at the current time. It has the following properties:
\begin{itemize}
    \item $\bx\now(.)$ is the minimizer of $$ \inf_{\bx\now} \left[ \int_\everything | \bq - \bx\now(\bq) |^2 \rho\init(\bq) \ d^2\bq \right]$$ subject to mass conservation (\ref{eq:OTcnstr});
    \item $\bx\now(.)$ is also the gradient of the (convex) Kantorovich potential $\cpot$; 
    \item $\cpot$ is the solution of the Monge-Amp\`ere equation (\ref{eq:MA});
    \item the convexity of $\cpot$ implies that there is no multi-streaming in the reconstructed dynamics;
    \item $\cpot$ is related to the gravitational potential at the initial condition $\gpot\init$ by:
    $$ \cpot(\bq) = 1/2 \bq^2 - \gpot\init(\bq).$$
\end{itemize}
From the assignment map $\bx\now(.)$, it is (optionally) possible to reconstruct higher-order dynamics using Lagrangian perturbation theory \citep{LagrangianPerturbTheory,1995MNRAS.276..115C} and at first order using the expression we have given in  (\ref{eq:secondorder}). \\

\section{solving the assignment problem}
\label{sec:assign}

In this section, we describe numerical solution mechanisms to compute the assignment map $\bx\now(.)$. We first review the existing methods, that are based on a 
discretization of the density $\rho\init$ at the initial condition and a discretization of the density $\rho\now$ at current time (\S\ref{sec:discreteMAK},\S\ref{sec:dualMAK}). Then we present our method 
(based on semi-discrete optimal transport), that uses a continuous representation of the initial density $\rho\init$ and a discrete representation of the 
density $\rho\now$, hence a \emph{semi-discrete} method (\S\ref{sec:sdMAK}).

\subsection{Discrete-discrete MAK reconstruction}
\label{sec:discreteMAK}

We consider (for now) that the density at the initial condition $\rho\init$ and the density at present time $\rho\now$ are both
represented in discrete form, by a set of $N$ particles. We consider that each particle $i$ has a mass $\mu_i = 1/N$:
\begin{itemize}
    \item At the initial condition $\tau\init$, the mass distribution $\rho\init$ is represented by a set of $N$ points $\bq_i, i=1\ldots N$. 
       Since the initial distribution of mass is uniform at $\tau\init$, the points $\bq_i$ are organized on a regular grid;
    \item at present time $\tau\now$, the distribution of mass $\rho\now$ is represented by a set of (the same number $N$) of points $\bx_j, j=1\ldots N$.  
\end{itemize}

In this setting, the initial problem of finding the (continuous) map $\bx\now(.): \everything \rightarrow \everything$ is replaced with finding which point $\bx_j$ corresponds
to each point $\bq_i$. The discrete version of Monge's problem (\ref{eq:OTcost}) writes:
\be
\label{eq:discreteMAK}
\inf_{\pi} \sum_j | \bq_j - \bx_{\pi(j)} |^2
\ee
where $\pi: [1 \ldots N] \rightarrow [1 \ldots N]$ is a permutation of the indices. \\

Note that the discrete Monge problem (\ref{eq:discreteMAK}) is purely combinatorial. Conceptually, one can imagine solving it by systematically testing the $N!$ possible permutations $\pi$. 
Clearly, it is not feasible in practice.  However, there exists more efficient algorithms  which  
guarantee  that  the  optimal assignment is found.
Faster assignment algorithms have been
developed with polynomial complexity \citep{Henon2002,Bertsekas1989}. 
The latest algorithm developed by M. H\'enon and used in
our previous works \citep{EUR}, which is a cosmological 
adaptation of the auction algorithm of Bertsekas, scales approximately as
$N^{3}$ (for relevant details see, e.g. \citep{Bertsekas1992}). Later improvements
of the auction algorithm allowed to make it faster (see \cite{OTReviewMerigotThibert}, Section 3).
However, even with these improvements, these combinatorial algorithms remain slow for all practical 
purposes and this has been a major obstacle
for the progress of MAK reconstruction in the past few years and since its first application to cosmology.
Such algorithms have proved too slow for the cosmological analyses of 
large datasets, or those that require repeated reconstructions. A notable example of such an instance is
the reconstruction of detailed features of the primordial density fluctuation
field or the primordial power-spectrum and in particular the reconstruction of the baryonic acoustic oscillations.
For a proper reconstruction of acoustic peaks one needs to treat extremely large datasets and/or 
carry reconstruction on a very large number of simulations for a proper handling of errors.

\subsection{MAK duality}
\label{sec:dualMAK}

We now exhibit more structure in the discrete Monge problem (\ref{eq:discreteMAK}), and its relation with the gravitational potential $\gpot\init$, that
we will use to design a more efficient algorithm. 

Instead of searching for the (combinatorial) assignment $i \mapsto j = \pi(i)$, 
we consider now the following optimization problem, introduced by \cite{Kanto:1942}:

\ba
\label{eq:OTplan}
\inf_\gamma \sum_i \sum_j \gamma_{ij} | \bx_i - \bq_j |^2 \quad \mbox{subject to} \\
\label{eq:OTplanC1}
\sum_i \gamma_{ij} = \mu_i \\
\label{eq:OTplanC2}
\sum_j \gamma_{ij} = \mu_j \\
\label{eq:OTplanC3}
\gamma_{ij} \ge 0 \quad \forall i,j. 
\ea

The objective function (\ref{eq:OTplan}) depends on an $[1,N]^2 \rightarrow \mathbb{R}^+$ array of coefficients $\gamma_{ij}$.
Intuitively, each coefficient $\gamma_{ij}$ indicates how much matter goes from $\bq_j$ to $\bx_i$. 
In this setting, matter can split and merge between different particles, for instance, a particle $\bq_j$ 
can send half of its matter to particle $\bx_k$ and the other half to particle $\bx_l$ (using $\gamma_{jk} = \gamma_{jl} = 0.5$). Clearly, 
the mass of all the matter that gathers at a particle $\bx_i$ should sum as the mass $\mu_i$ of the particle (constraint (\ref{eq:OTplanC1})), and 
the mass of all the matter originated from a particle $\bq_j$ should sum as the mass $\mu_j$ of the particle (constraint (\ref{eq:OTplanC2})). Since 
no matter disappears, all coefficients $\gamma_{ij}$ should be positive (constraint (\ref{eq:OTplanC3})). An array of coefficients $\gamma_{ij}$ that 
satisfies the three constraints is called a \emph{transport plan} (and an \emph{optimal} transport plan if it minimizes (\ref{eq:OTplan})). \\

At first sight, it may seem to be a rather convoluted re-formulation of Monge's problem, in particular, we now need to find $N^2$ unknowns, to be compared with
the $N\rightarrow N$ permutation we had to find initially. However, it can be observed that $(\ref{eq:OTplan})$ is a \emph{linear} optimization problem with \emph{linear} constraints.
Introduce $\dgpot \in \mathbb{R}^N$ and $\gpot \in \mathbb{R}^N$ the Lagrange multipliers associated with constraints (\ref{eq:OTplanC1}) and (\ref{eq:OTplanC2}) 
respectively (note that we use the same notation $\dgpot$ for the Lagrange multiplier of the constraint (\ref{eq:OTplanC2}) and the gravitational 
potential, we elaborate on that in the next subsection). The dual of the optimization problem (\ref{eq:OTplan}) writes (see the tutorials in \cite{OTReviewMerigotThibert,OTON,Santambrogio,DBLP:journals/cg/LevyS18} and the references herein):

\ba 
\label{eq:DDK}
&& \sup_{\dgpot,\gpot} \left[ \sum_i \dgpot_i \mu_i + \sum_j \gpot_j \mu_j \right]\nonumber\\
\label{eq:DDKcnstr}
&& \mbox{subject to } \dgpot_i + \gpot_j \le 1/2 | \bx_i - \bq_j |^2 \quad \forall i,j.
 \nonumber\\
\ea

In addition, given a pair $\dgpot, \gpot$ that satisfies the constraint (\ref{eq:DDKcnstr}), it is easy to check that replacing $\gpot$
with $\dgpot^c$ still satisfies the constraints while always increasing the objective function (\ref{eq:DDK}), where $\dgpot^c$ is defined by:
\be
   \dgpot^c_j = \inf_i \left[ 1/2 |\bx_i - \bq_j|^2 - \dgpot_i \right].
   \label{eq:cconj}
\ee
 
There exists several methods that exploit the structure of the problem (\ref{eq:OTplan}) and its dual (\ref{eq:DDK}), we refer the reader to 
\citet{ComputationalOT,Santambrogio} for a survey. Among these methods, we mention the \emph{entropic regularized} method, that solves:
\ba
\label{eq:regOTplan}
\nonumber
\inf_\gamma \left[\sum_i \sum_j \gamma_{ij} | \bx_i - \bq_j |^2 + \epsilon \sum_i\sum_j \gamma_{ij} \log(\gamma_{ij}) \right] \\
   \mbox{subject to } (\ref{eq:OTplanC1})\, (\ref{eq:OTplanC2})\, (\ref{eq:OTplanC3})
\ea
where $\epsilon$ is a (small) regularization parameter. 
If both the $\bq_j$'s and $\bx_i$'s are supported by regular grids, it is possible to exploit the structure of (\ref{eq:regOTplan}) to design a fast and efficient algorithm \citep{DBLP:conf/nips/Cuturi13}.
The advantage of this algorithm is its speed and simplicity. The drawbacks are the need for re-sampling everything on regular grids and the difficulty of tuning the parameter $\epsilon$
(a too large value of $\epsilon$ results in a blurry, imprecise transport plan, and a too small value of $\epsilon$ makes the algorithm slow to converge). Various ways of leveraging its speed while overcoming its drawbacks are currently being studied \citep{BenamouPriv}. 

In the next subsection we describe a different method and although the algorithm that we eventually obtain is more complicated than those based on entropic regularized schemes, it does not depend a regularization parameter $\epsilon$, and does not 
require $\rho\now$ to be re-sampled on a regular grid.

\begin{figure}
    \centerline{
	   \includegraphics[width=0.5\columnwidth]{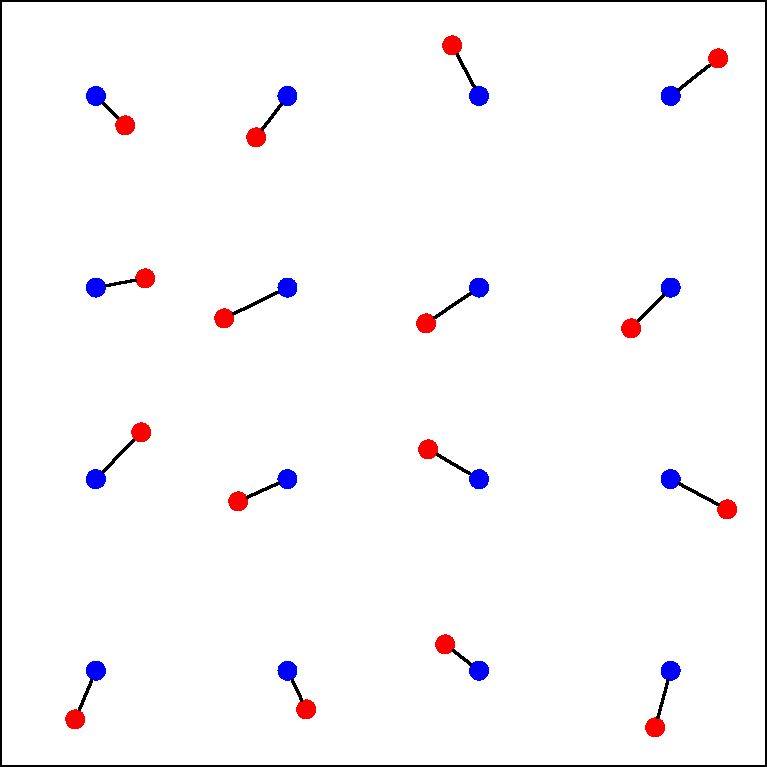}
       \includegraphics[width=0.5\columnwidth]{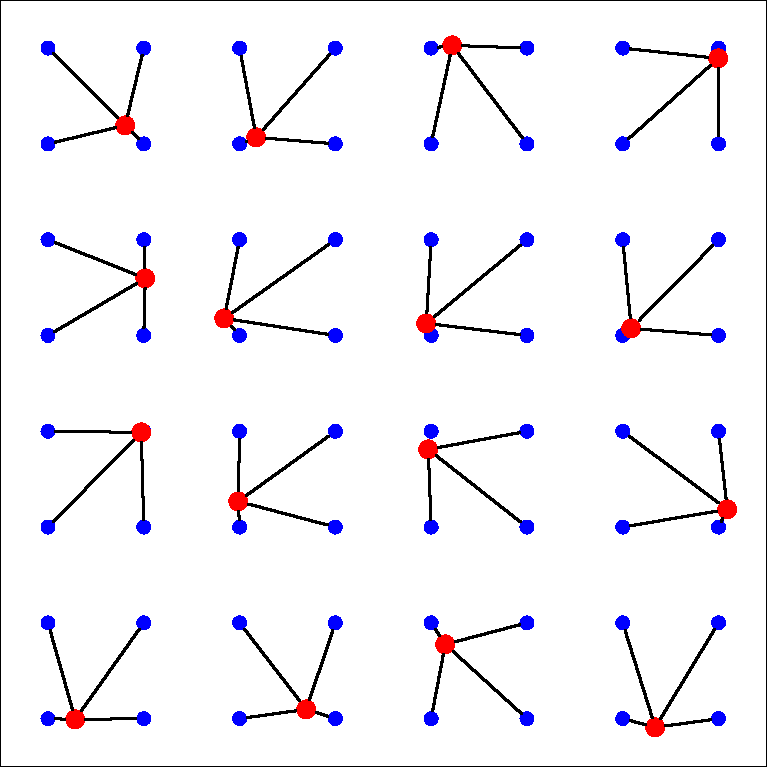}
    }
    \centerline{A \hspace{0.5\columnwidth} B}
    \vspace{3mm}
    \centerline{
	   \includegraphics[width=0.5\columnwidth]{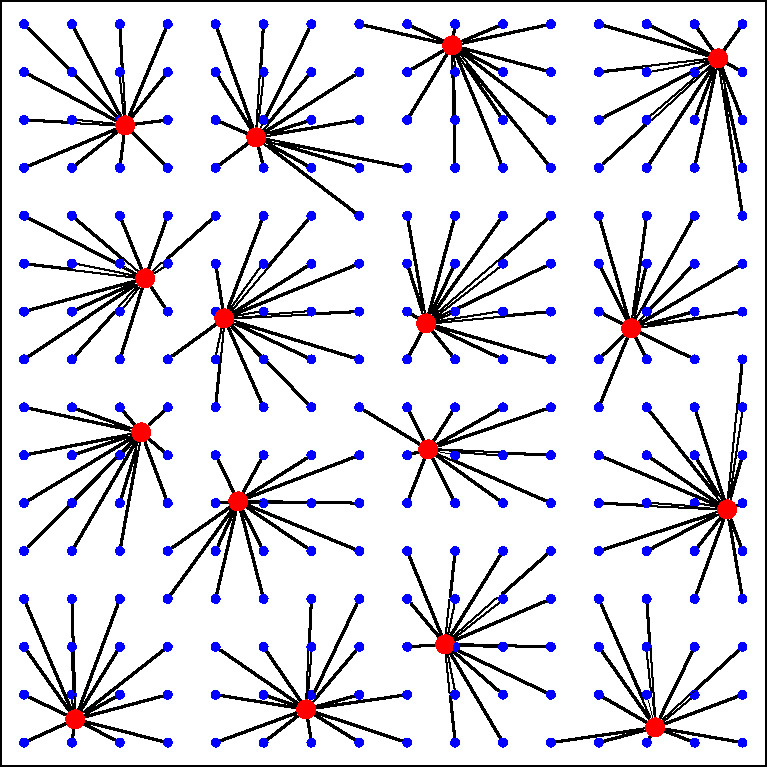}
       \includegraphics[width=0.5\columnwidth]{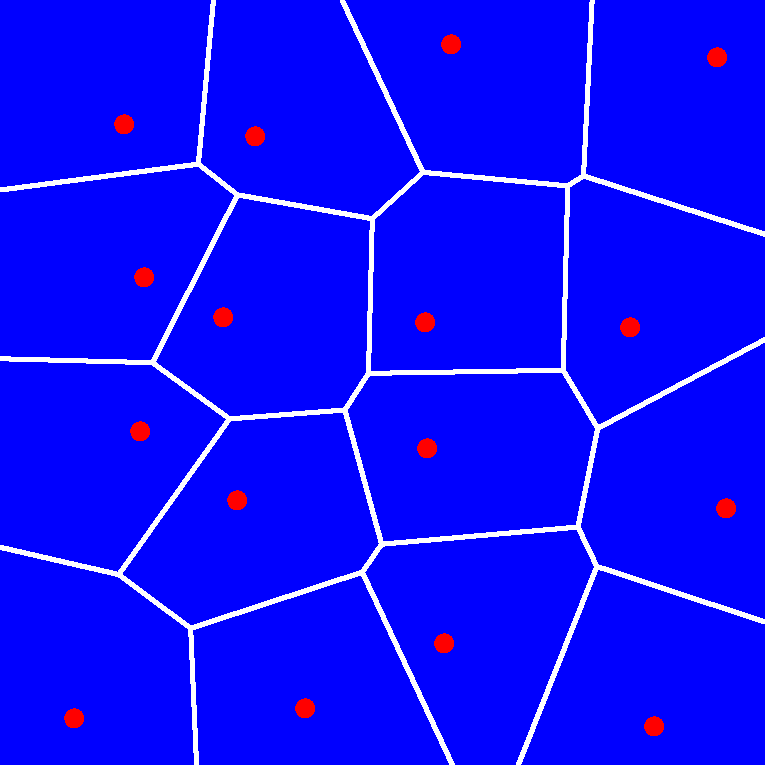}
    }
    \centerline{C \hspace{0.5\columnwidth} D}
    \caption{From discrete-discrete to semi-discrete transport. A: coupling between the $\bx_i$ points (in red) and the $\bq_j$ points (in blue). B: each red $\bx_i$ point is coupled with 4 $\bq_j$ points, each of them with 1/4 the mass. C: each red $\bx_i$ point is coupled with 16 $\bq_j$ points. D: at the limit, when the number of $\bq_j$ points tends to infinity, it can be proved that the optimal assignment couples each $\bx_i$ point with a polygonal area (called a Laguerre cell).}
    \label{fig:transport}
\end{figure}

\subsection{Semi-discrete MAK reconstruction}

\label{sec:sdMAK}

Consider the discrete assignment problem expressed by (\ref{eq:discreteMAK}). The density $\rho\init$ is represented by a set of $N$ particles $(\bq_j)_{j=1}^N$ located on a regular grid (in blue in Figure \ref{fig:transport}-A). The distribution of mass $\rho\now$ at current time is a set of $N$ particles $(\bx_i)_{i=1}^N$ (in red in the figure). Suppose one wishes to increase the precision by using a finer grid for the $\bq_j$'s. For instance, in Figure \ref{fig:transport}-B, 4 points $\bq_j$ are coupled to each point $\bx_i$. Up to now we supposed that there was the same number of points on both sides, but one may imagine that each $\bx_i$ is replaced by four points located at the same position with 1/4 the mass allocated to each of them. We can further refine the grid, as shown in 
Figure \ref{fig:transport}-C, where 16 points $\bq_j$ are coupled to each point $\bx_i$. Clearly, doing so will significantly increase computation time, but consider now that the number of $\bq_j$ particles tends to infinity, while each particle's mass tends to zero accordingly. At the limit, $\rho\init$ tends to the uniform density, while $\rho\now$ is still supported by the set of points $(\bx_i)$. 

The above setting corresponds exactly to our early universe reconstruction problem, where the initial density $\rho\init$ is uniform, and the density at present time $\rho\now$ 
corresponds to a set of galaxies, each of them represented by a single point $\bx_i$. In this setting, as detailed below, it can be proved that each point $\bx_i$ is coupled to a polyhedral region (or polygonal in 2D, see Figure \ref{fig:transport}-D), that can be computed explicitly by an algorithm. Not only the so-computed result is more precise, but also the algorithm is much faster than the combinatorial one used in Figure \ref{fig:transport}-A,B,C. 
 
\begin{figure}
    \centerline{
	   \includegraphics[width=0.5\columnwidth]{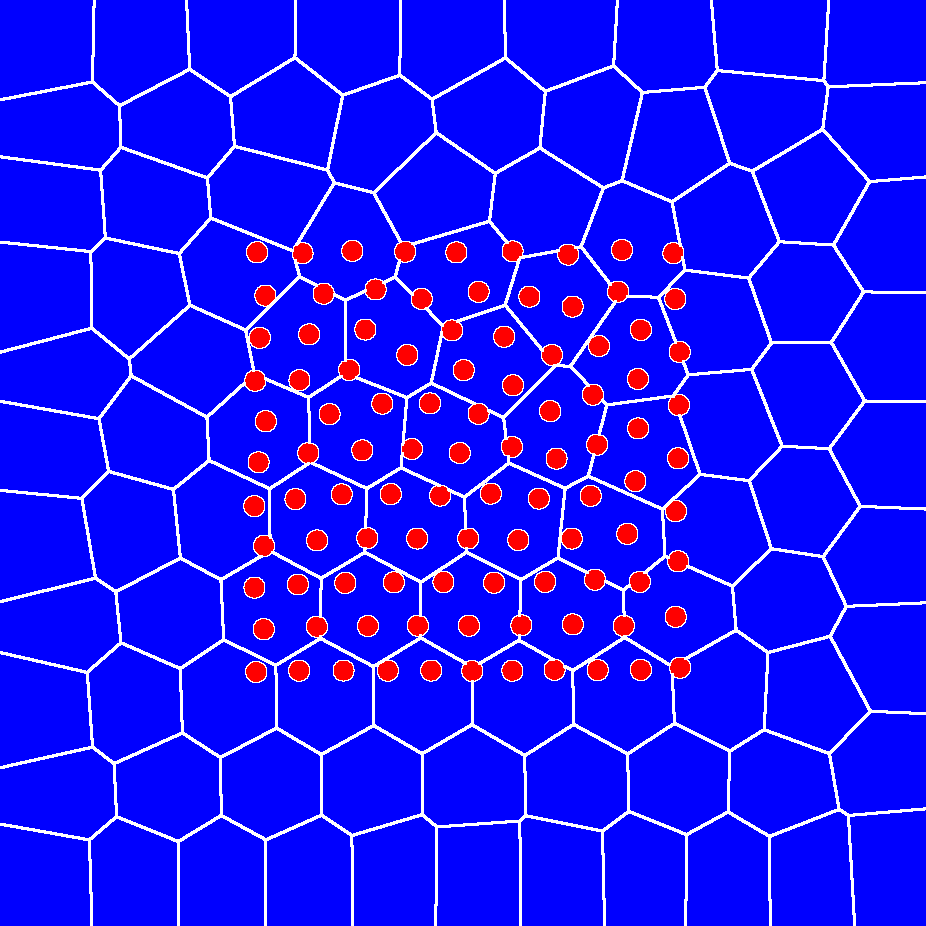}
       \includegraphics[width=0.5\columnwidth]{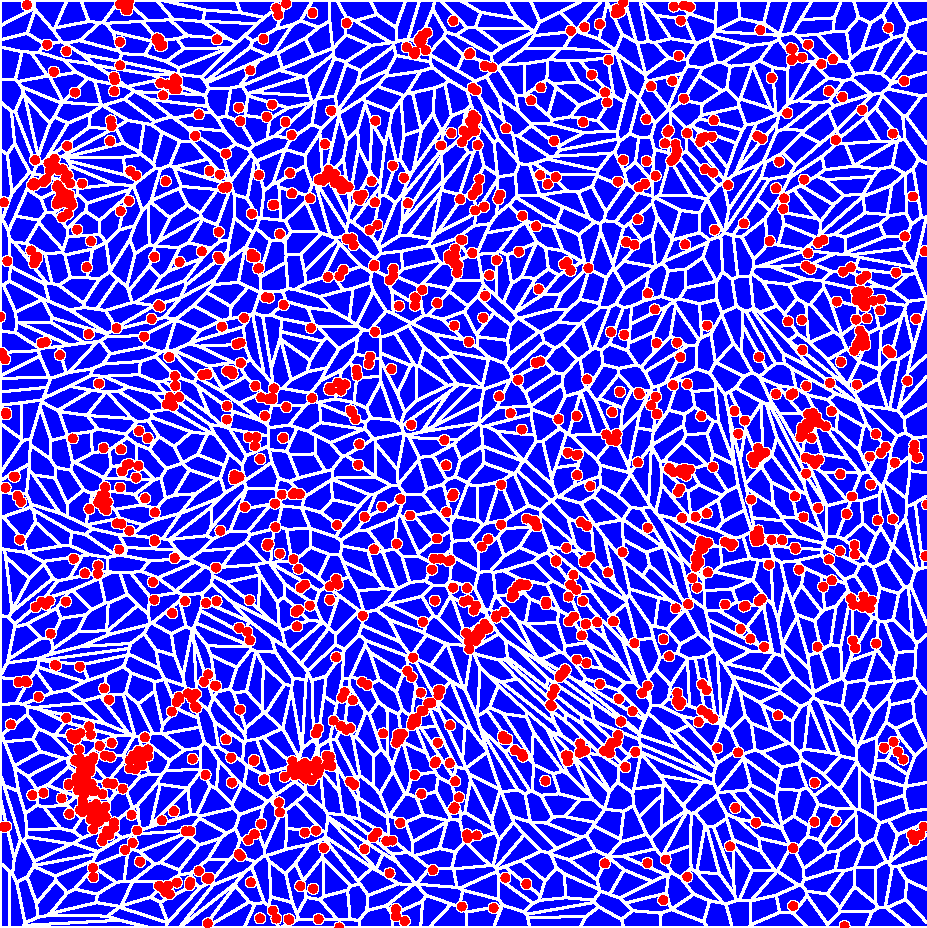}
    }
    \centerline{A \hspace{0.5\columnwidth} B}
    \vspace{3mm}
    \centerline{
	   \includegraphics[width=0.5\columnwidth]{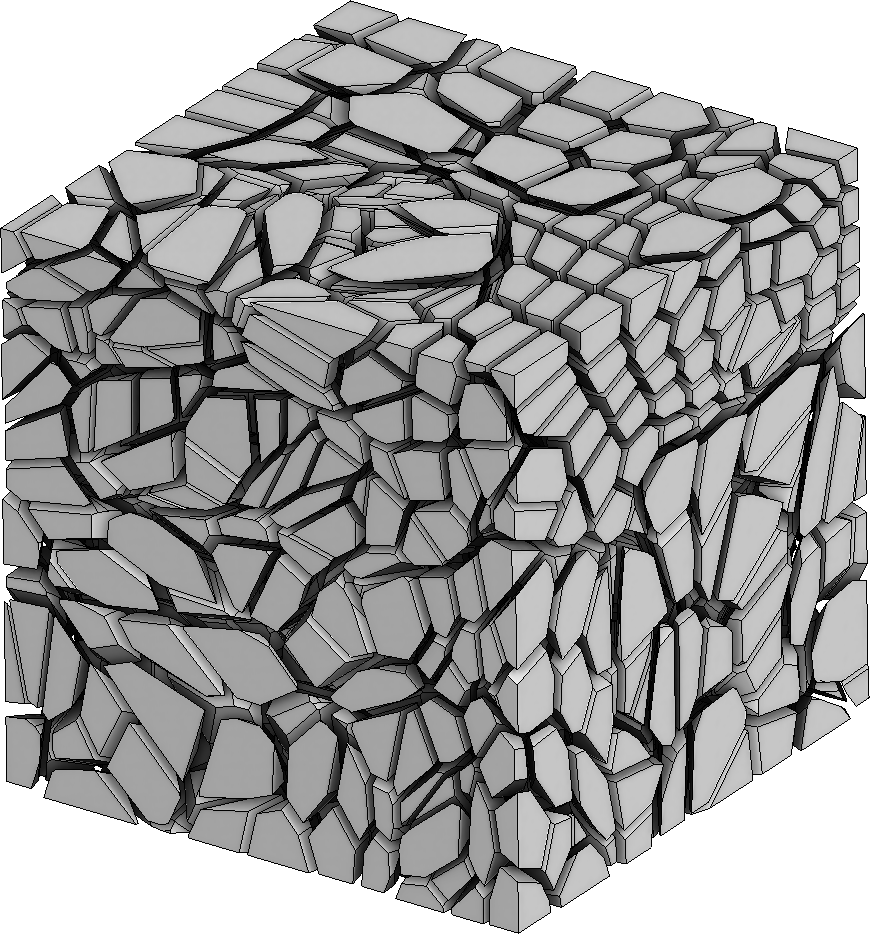}
       \includegraphics[width=0.5\columnwidth]{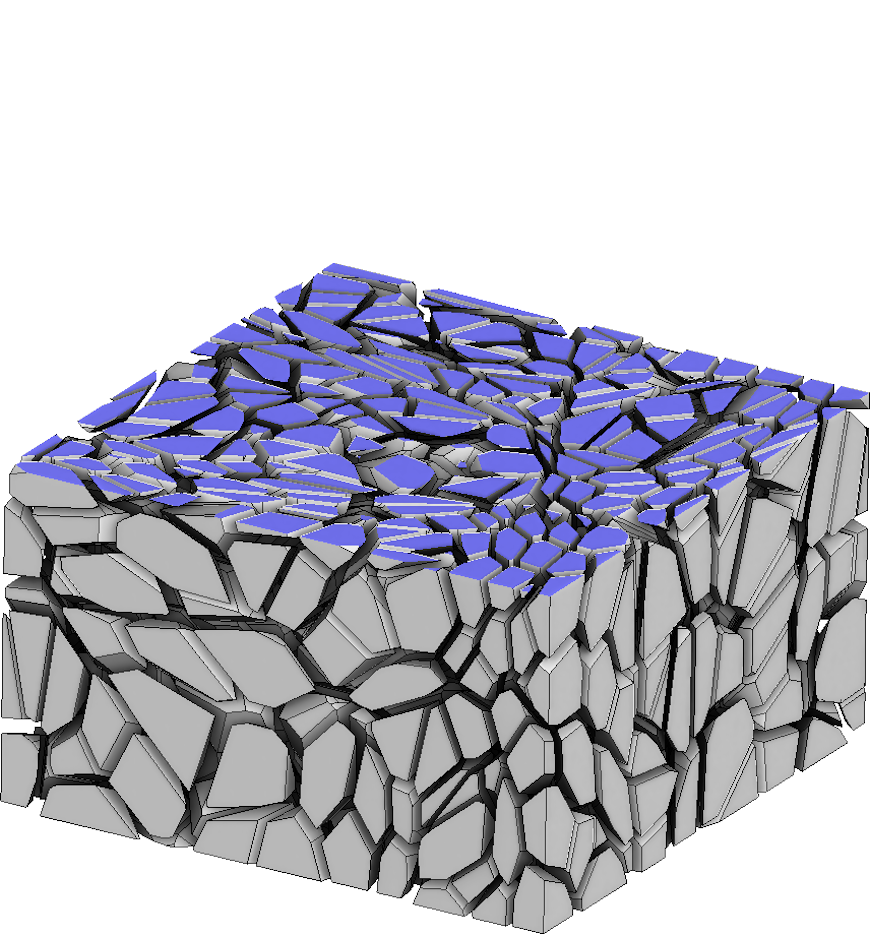}
    }
    \centerline{C \hspace{0.5\columnwidth} D}
    \caption{A: the Laguerre cell $V_i^\psi$ associated with the point $\bx_i$ does not necessarily contain $\bx_i$. B: a 2D Laguerre diagram with 1500 cells. 
             C,D: a 3D Laguerre diagram with 1500 cells and a cross-section (cells are slightly shrunk to ease visualization).}
    \label{fig:Laguerre}
\end{figure}

Next, We give more details about these polyhedral regions by first 
considering the dual problem in expression (\ref{eq:DDK}). The relation between the primal and dual problem that we wrote in the discrete-discrete setting remains valid in our continuous-discrete case (with integrals instead of discrete sums), and the dual problem writes:

\ba
\nonumber
\sup_{\gpot,\dgpot} \left[ \int_\everything \gpot(\bq) \rho\init(\bq) \ d^3\bq + \int_\everything \dgpot(\bx) \rho\now(\bx) \ d^3\bx \right] \\ \mbox{ subject to}\quad  \gpot(\bq) + \dgpot(\bx) \le 1/2 | \bx - \bq |^2 \quad \forall \bx, \bq. 
\ea

\noindent
We recall that $\rho\now$ is discrete, supported by the $\bx_i$'s. Then the Lagrange multiplier $\psi$ is determined by the vector $\psi_i = \psi(\bx_i)$, and the dual problem becomes:
\ba
\sup_{\gpot,\dgpot} \left[ \int_\everything \gpot(\bq) \rho\init(\bq) \ d^3\bq + \sum_i \psi_i \mu_i \right] \nonumber \\ \mbox{ subject to}\quad  \gpot(\bq) + \dgpot_i \le 1/2 | \bx_i - \bq |^2 \quad \forall \bx_i, \bq 
\ea

The optimization problem depends on a vector of $N$ values $(\dgpot_i)$ and a function $\gpot: \everything \rightarrow \mathbb{R}$.
As we mentioned, replacement of $\gpot$ by $\dgpot^c$ defined in (\ref{eq:cconj}) always increases the objective function, thus we can now deduce $\gpot = \dgpot^c$ from $\dgpot$ 
and consider the following optimization problem:
\ba
\label{eq:ddk1}
\sup_\psi \left[ \int_\everything \inf_i \left[ 1/2| \bx_i  - \bq |^2 - \psi_i \right] \rho\init(\bq) d^3\bq + \sum_i \psi_i \mu_i \right] \nonumber\\
\mbox{subject to } \dgpot^{cc} = \dgpot 
\label{eq:Ccconvex}
\ea 
that solely depends on a vector $\dgpot$ of $N$ components. 
The constraint (\ref{eq:Ccconvex}) ensures that $\dgpot$ is such that there can exist at least one pair $\gpot,\dgpot$ that satisfies (\ref{eq:DDKcnstr}). It also means that the associated Kantorovich potentials $\cpot,\dcpot$ (\S\ref{sec:Kpot}) are convex.

Consider now the integrand of (\ref{eq:ddk1}). It is possible to partition $\everything$ into the set of $N$ regions $(\everything^\psi_i)$, defined according to the index $i$ that realizes the infimum in the integrand of (\ref{eq:ddk1}). This lets
rearrange the integral as follows:

\ba
\label{eq:SDISC}
\sup_\psi \left[ \sum_i \int_{\everything^\psi_i} 1/2\left[|\bx_i - \bq|^2 - \psi_i \right]\rho\init(\bq) d^3\bq + \sum_i \psi_i \mu_i \right] \nonumber\\
\mbox{subject to } \everything_i^\psi \neq \emptyset \quad \forall i 
\label{eq:Ccconvex2}
\ea
where
\ba
\nonumber \everything^\psi_i = \left\{ \bq \ | \ 1/2|\bx_i - \bq|^2 - \psi_i < 1/2|\bx_j - \bq|^2 - \psi_j \quad \forall j \neq i \right\} \\ 
\ea

The so-defined partition of $\everything$ into the regions $\everything^\psi_i$ is called a \emph{Laguerre diagram} (or power diagram in our specific case). An individual region is called a \emph{Laguerre cell}. Laguerre diagram is a generalization of 
\emph{Voronoi diagram}, parameterized by the vector $\psi \in \mathbb{R}^N$. If $\psi_i = 0$ for all $i$, then the Laguerre diagram is a Voronoi diagram. Voronoi diagram is used in existing cosmological ALE codes such as \cite{DBLP:journals/cse/WhiteS99}. In our setting, the additional vector $\psi$, that corresponds to the gravitational potential, makes it possible to control the volumes of the Laguerre cells through the semi-discrete Monge-Ampère equation. 
Examples of Laguerre diagrams in 2D and 3D are shown in Figure \ref{fig:Laguerre}. The Laguerre diagram is completely defined from the points $(\bx_i)_{i=1}^N$ and the vector $\psi \in \mathbb{R}^N$ of coefficients. Note that in contrast with a Voronoi cell, depending on the vector $\psi$, the Laguerre cell $V_i^\psi$ associated with a point $\bx_i$ does not necessarily contain $\bx_i$ (see Figure \ref{fig:Laguerre}-A)). It is even possible for a cell $V_i^\psi$ to be empty. The examples shown in Figure \ref{fig:Laguerre} are solutions of the optimal transport problem, thus all cells have the same area/volume. They have different shapes though, this is because the MA equation is non-linear, with potentially a highly anisotropic solution. This anisotropy influences the shapes of the Laguerre cells. 

In terms of the Laguerre diagram, the convexity constraint (\ref{eq:Ccconvex}) means that no Laguerre cell is empty (\ref{eq:Ccconvex2}). 
The notion of Laguerre diagram is well known and well studied in computational geometry \citep{DBLP:journals/siamcomp/Aurenhammer87}, this equips us 
with efficient computational tools to solve the optimization problem (\ref{eq:SDISC}), as explained in the next section.

The solution to the optimization problem (\ref{eq:SDISC}) results in a vector $\dgpot$ of $N$ coefficients, from which one can subsequently deduce the gravitational potential $\gpot\init$ using the relation $\gpot\init = \dgpot^c$, or:
\ba
   \gpot\init(\bq) & = & \inf_i \left[ 1/2 | \bq - \bx_i |^2 - \psi_i \right] \nonumber \\
                   & = & 1/2 | \bq - \bx_{i(\bq)} |^2 - \psi_{i(\bq)} \,\,\, ,
\ea
where $i(\bq)$ is the index of the Laguerre cell $\everything_i$ that contains $\bq$. Note that all the $\bq$'s located in the Laguerre 
cell $\everything^\psi_i$ are mapped to $\bx_i$ through $\bx\now(.)$:
\ba
\bx\now(\bq) & = & \bq - \nabla_q \gpot\init(\bq) \nonumber \\
             & = & \bq - \nabla_q \left( 1/2|\bq - \bx_{i(\bq)}|^2 - \dgpot_{i(\bq)} \right) \nonumber \\
             & = & \bx_{i(\bq)}.  
\ea             

To summarize, the solution of the optimization problem (\ref{eq:SDISC}) gives a vector $(\dgpot_i)$ of $N$ coefficients. These coefficients
define a partition of $\everything$ into $N$ Laguerre cells $(\everything^\dgpot_i)$. Each Laguerre cell $\everything^\psi_i$ corresponds to the (continuous) 
set of points $\bq$ at the initial condition that collapses into a given point $\bx_i$ at current time. In other words, the Laguerre cell
$\everything^\psi_i$ corresponds to the pre-image of $\bx_i$ through $\bx\now$.

\begin{figure*}
    \centerline{
	   \includegraphics[width=0.52\columnwidth]{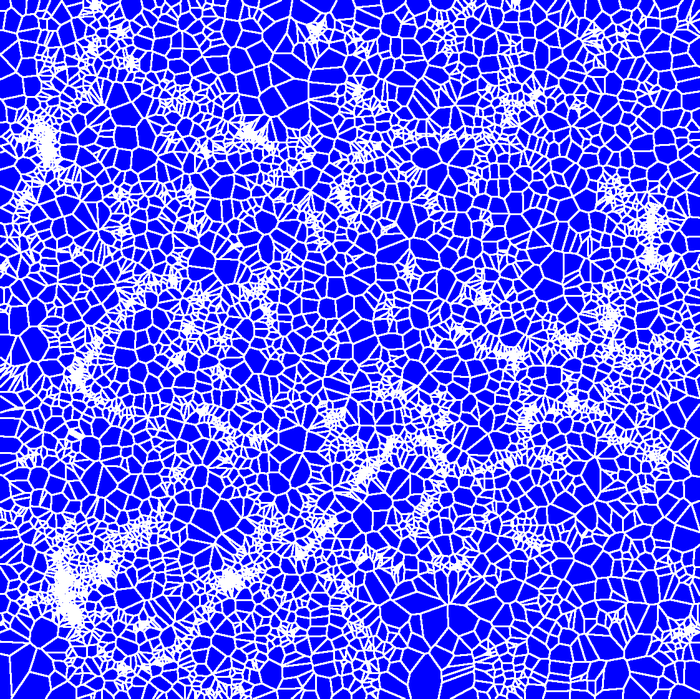}
       \includegraphics[width=0.52\columnwidth]{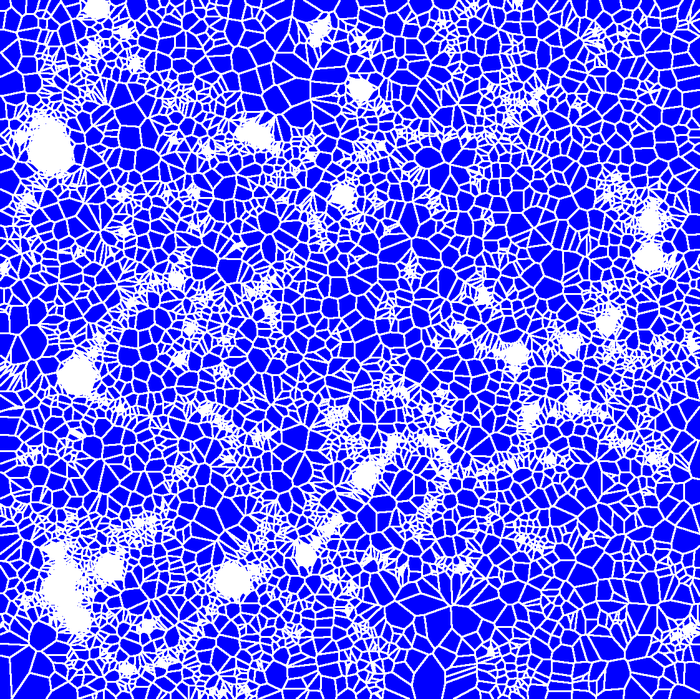}
 	   \includegraphics[width=0.52\columnwidth]{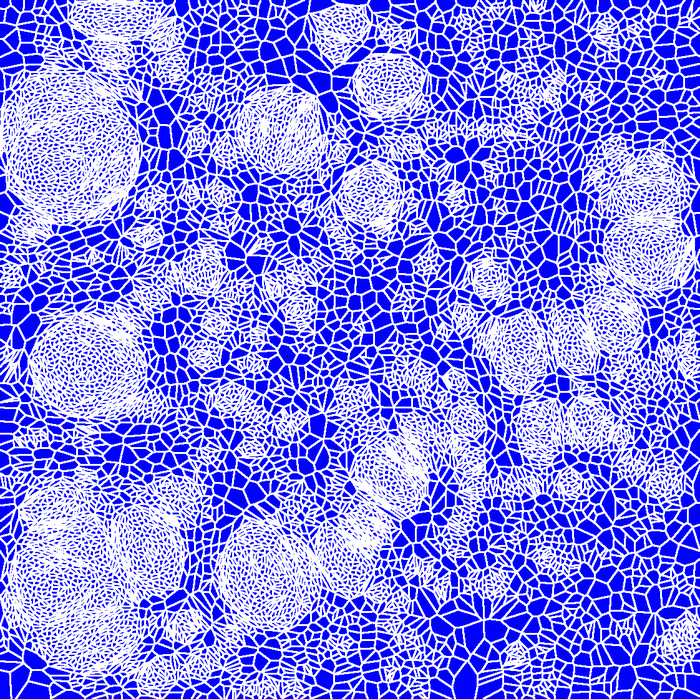}
       \includegraphics[width=0.52\columnwidth]{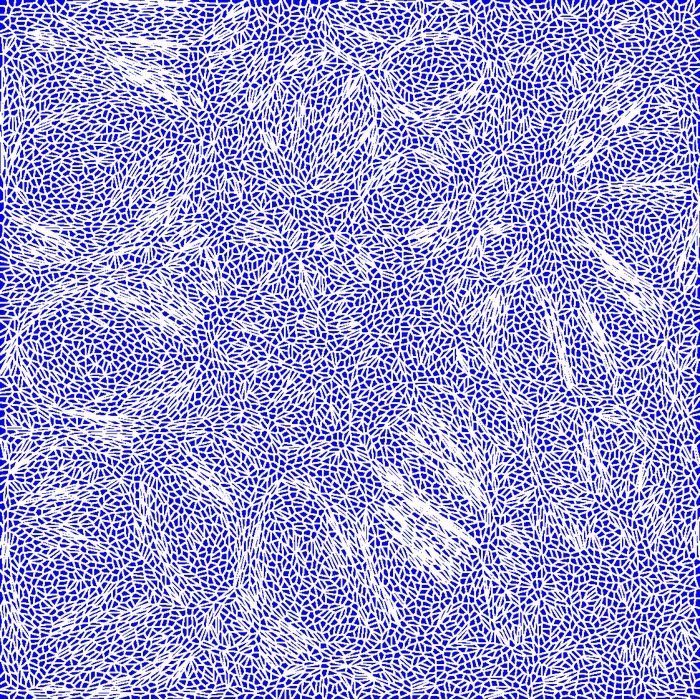}
    }
    \caption{Newton's algorithm applied to a 2D problem with 10000 points. From left to right: iterations 0, 4, 8 and 12. At iteration 12, all cells have the same area (largest cell area error is smaller than 1\%). The diagram is the unique optimal transport solution.}
    \label{fig:Newton}
\end{figure*}

\section{Numerical solution mechanism}
\label{sec:num}

Let us denote by $K(\psi): \mathbb{R}^N \rightarrow \mathbb{R}$ the objective function of the optimization problem (\ref{eq:SDISC}):
\be
   K(\psi) = \sum_i \int_{\everything^\psi_i} 1/2|\bx_i - \bq|^2 - \psi_i \rho\init(\bq) d^3\bq + \sum_i \psi_i \mu_i  
\ee
It can be shown that $K(.)$ is a concave $C^2$ function, which suggests that it can be efficiently maximized by a Newton algorithm \citep{DBLP:journals/corr/KitagawaMT16,DBLP:conf/compgeom/AurenhammerHA92,journals/M2AN/LevyNAL15}. 
The extensive algorithmic details of this Newton algorithm are given in Appendix \ref{sec:algo}. In a nutshell, the algorithm iteratively maximizes second-order approximations of $K(.)$. A 2-dimensional example of the Laguerre diagrams corresponding to each iteration is shown in Figure \ref{fig:Newton}. The algorithm starts with $\psi = 0$, then updates $\psi$ by solving a series of linear system. In the end, the algorithm finds the unique solution, and all the Laguerre cells have the prescribed volumes. This \emph{numerical} algorithm outperforms the previous \emph{combinatorial} ones, by fully exploiting the variational nature of the problem. 

\subsubsection*{Performance}

As can be seen in Fig.~\ref{fig:complexity} the computation time indeed scales as $\mathcal{O}\left( N\log N\right)$, massively outperforming previous approaches.  To provide a realistic setting within which we aim to use this algorithm, we employ snapshots from the cosmological $N$-body simulation suite \textsc{AbacusCosmos}~\citep{Garrison2018} (see next section).  The convergence time of the reconstruction among others slightly depends on the degree to which non-linear clustering has occurred in the samples.  We therefore run the complexity analysis on snapshots of different redshifts, where per sample size $ N$, the particles are kept across redshifts, and find that the computation time increases with decreasing redshift.  
\begin{figure}
    \centering
	\includegraphics[width=\columnwidth]{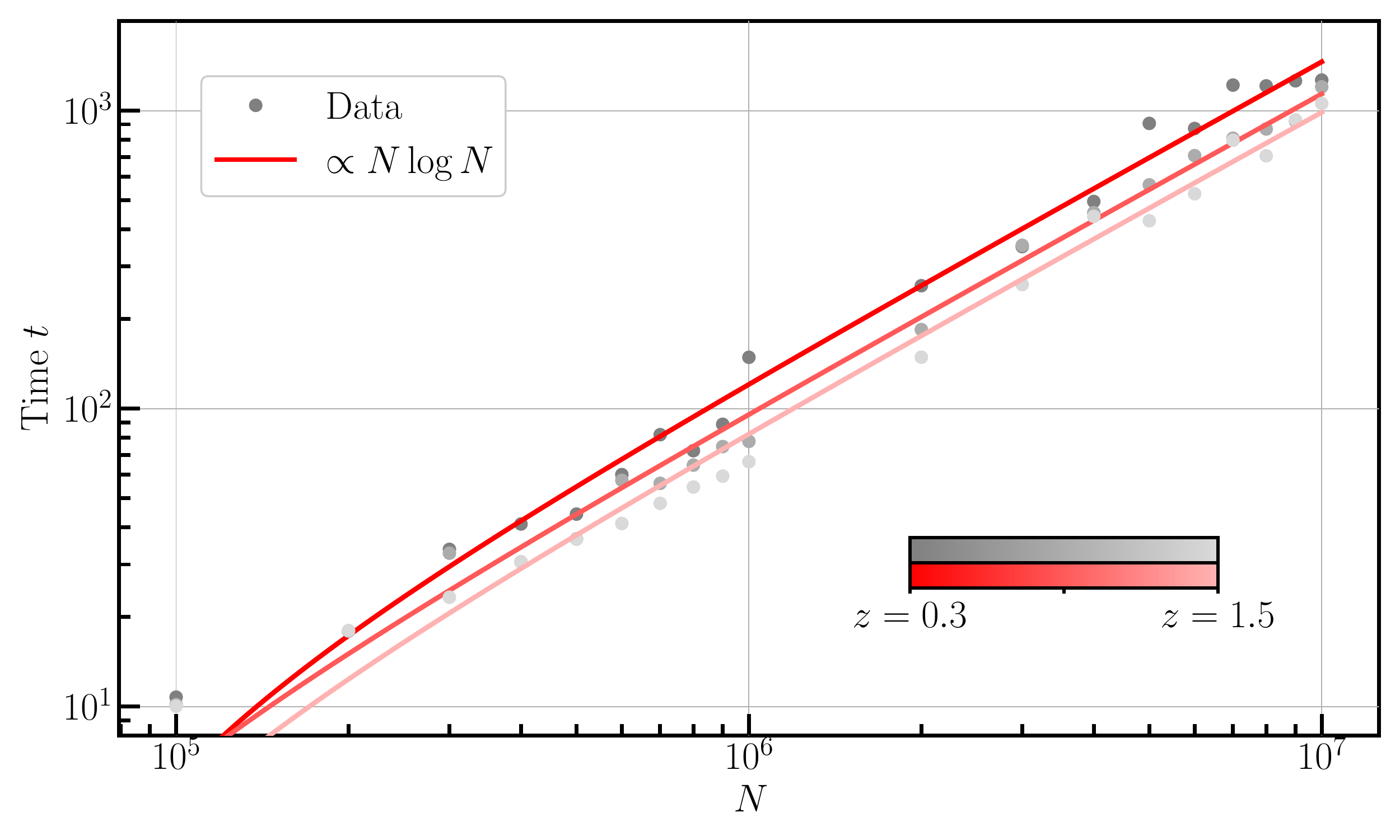}
    \caption{Empirical computational complexity of the semi-discrete algorithm, running on an Intel Xeon 5122 3.6 GHz with 128 Gb RAM and an NVidia V100 GPU with 16 Gb RAM. Tests were done on \textsc{AbacusCosmos} simulations.  The reconstructions were performed on particle samples of different sizes at redshifts of $z_s=0.3$, $0.7$, and $1.5$, resulting in different computation times ({\it gray dots}).  Due to increase of non-linearities towards lower redshifts, the reconstructions generally perform faster at higher redshifts.  This is shown by the data points, and emphasised by analytical fits to those points ({\it red lines}).}
    \label{fig:complexity}
\end{figure}

\section{Tests of the semi-discrete algorithm with cosmological simulations}
\label{sec:results}

Finally, we put our algorithm to the proof.  We employ a set of cosmological 
$N$-body simulations, in specific, 10 simulations} from the \textsc{AbacusCosmos} suite~\citep{Garrison2018,Garrison2019}\footnote{\url{https://lgarrison.github.io/AbacusCosmos/}}.  We present results of both qualitative and quantitative measures, which capture the basic capability of our code, and further highlight its special features.

\subsubsection*{The simulations}

The main field of application for reconstruction algorithms is the recovery of the linearly perturbed density field, e.g.~for improving the precision with which BAO can be measured.  At the same time, we aim to test our algorithm on smaller scales, where high resolution simulations are required.  We select ten ``AbacusCosmos\_1100box\_planck'' simulations from the \textsc{AbacusCosmos} suite, which are highly resolved large-scale simulations for $\Lambda$CDM cosmologies with parameters fixed to those of~\citet{Ade:2015xua}.  Each of the ten simulations has a box size of $(1100\,h^{-1}$Mpc$)^3$ and, for most of what follows, we sample $256^3$ of the $1440^3$ particles with which each simulation was run.  For later visualization and for the computation of power spectra we paint the particle positions onto a mesh of size $512^3$ by use of the python package \textsc{nbodykit}\footnote{\url{https://github.com/bccp/nbodykit/}}~\citep{Hand:2017pqn}.

While snapshots of the simulations are provided at a range of low redshifts, namely $z=$0.3, 0.5, 0.7, 1.0, and 1.5, we additionally generate the density field corresponding to their initial conditions at $z=49.0$ via \textsc{zeldovichPLT}\footnote{\url{https://github.com/lgarrison/zeldovich-PLT}}~\citep{Garrison:2016vvp}.  Also the density field is cast onto a mesh of size $512^3$.

\subsubsection*{The reconstructions}

We begin by computing the Laguerre diagram of a simulation's snapshot at a given redshift, $z_s$, as described in section~\ref{sec:num}. This assigns to each point $\bx_i$ the Laguerre cell $V_i^\psi$. Recall that each Laguerre cell $V_i^\psi$ represents the set of mass elements at initial positions $\bq$ in Lagrangian coordinates arriving at a given point $\bx_i$. The first-order Lagrangian approximation for the motion of each mass element as a function
of the redshift $z_f$, {\it i.e.} the Zel'dovich approximation, is given by Eq.~(\ref{eq:1D}) which we rephrase here for convenience:

\be
    \mathbf{x}(z_f) = \mathbf{q} + \frac{D(z_f)}{D(z_s)}(\mathbf{x}(z_s)-\mathbf{q}),
   \label{eq:zeldovichapprox2}
\ee
where $D(z)$ are the linear growth factors at redshifts $z$, which in our case are taken from the \textsc{AbacusCosmos} simulations themselves.\footnote{In practice, when the exact growth factors are not known, it suffices to use approximate values for a fiducial cosmology, from {e.g.} \citet{Lukic:2007fc}, and to relate the resulting amplitudes of the reconstructed linear densities to the correct amplitudes by a constant bias, {cf.} section~\ref{sub:cosmological quantities}.}

To this effect, each cell $V_i^\psi$ at the initial condition is shrunk towards a single point $\bx_i$ at the current time by interpolation, such that each mass element $\bq$ undergoes the motion governed by Eq.~(\ref{eq:zeldovichapprox2}). In order to analyse the reconstructed density field ({\it e.g.}~by computing its correlation function or power spectrum), we need to compute the Fourier transform of the corresponding density field. While it is possible to compute the Fourier transform of a Laguerre diagram \citep{FourrierVoro}, it is computationally very expensive. Hence, we simply convert our Lagrangian representation (Laguerre cells) into an Eulerian one (regular grid), as explained in Appendix \ref{sec:LagrangeEuler}. Once the density is represented in Euler form, we can use standard FFT-based analysis tools. 

In the following subsections we mainly focus on the computation of the initial positions of $256^3$ particles from two of the available redshifts --- the lowest available redshift, $z_s=0.3$, and the highest available, $z_s=1.5$ --- that can be seen as representative for low- and high-redshift samples of present galaxy surveys.  We estimate the algorithm's accuracy by reconstructing the density at $z_f=49$ in order to compare with the linear density corresponding to each of the simulations.  We present a non-exhaustive range of tests concerning the accuracy of the reconstruction algorithm as well as the ability to extract cosmological information from the obtained reconstructions.

\subsection{Qualitative diagnostics}
\label{sub:accuracy of the reconstruction}

\subsubsection*{Density slices}  

\begin{figure*}
\centering
\includegraphics[width=\textwidth]{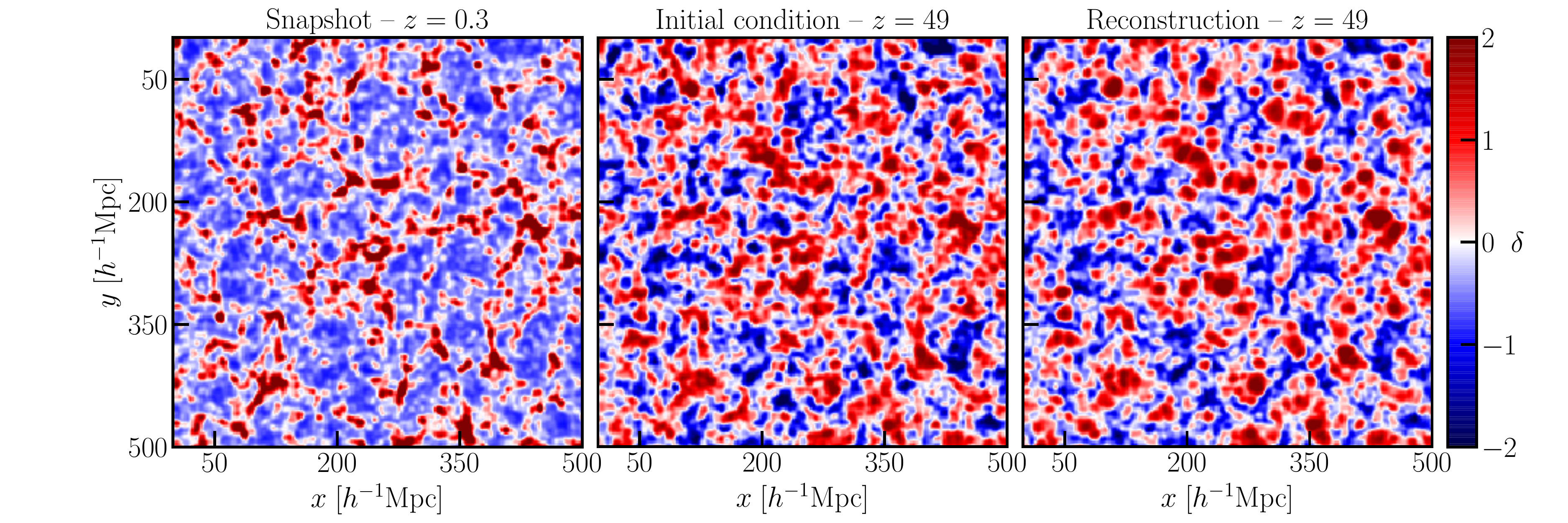}
\caption{Slices of snapshots and reconstruction of an \textsc{AbacusCosmos} simulation.  The panels show the same $(10\times500\times500)\,h^{-3}\rm{Mpc}^3$ slice of the density contrast, smoothed with a $2\,h^{-1}\rm{Mpc}$ Gaussian kernel to remove small-scale noise in the reconstruction.  All slices have been scaled to match the linear growth amplitude at $z=0.3$.}
\label{fig:threedensity}
\end{figure*}

Firstly, we present a purely visual comparison between the reconstructed and true density at $z_f=49$ by the example of a single simulation.  Figure~\ref{fig:threedensity} shows the density contrast as computed on a $512^3$-cell mesh averaged over a slice roughly of dimensions ($500\times500\times10$)$h^{-1}$Mpc.  Each slice was further smoothed with a $S = 3\,h^{-1}\rm{Mpc}$ Gaussian filter\footnote{The smoothing filter used here is defined as $W_S(k)=\exp\left[-k^2S^2\right]$.} to remove noise on the smallest scales.  The left panel shows the slice in the original, $z_s=0.3$, snapshot that exhibits strongly pronounced over-dense regions.  The center and right panels show the same slice in the $z_f=49$ initial condition and the reconstructed density field, respectively, and are nearly indistinguishable.  This rather qualitative comparison already anticipates corresponding agreement between more quantitative measures that are presented in the following paragraphs.

\subsubsection*{1-point distributions}

\begin{figure*}
\centering
\includegraphics[width=\textwidth]{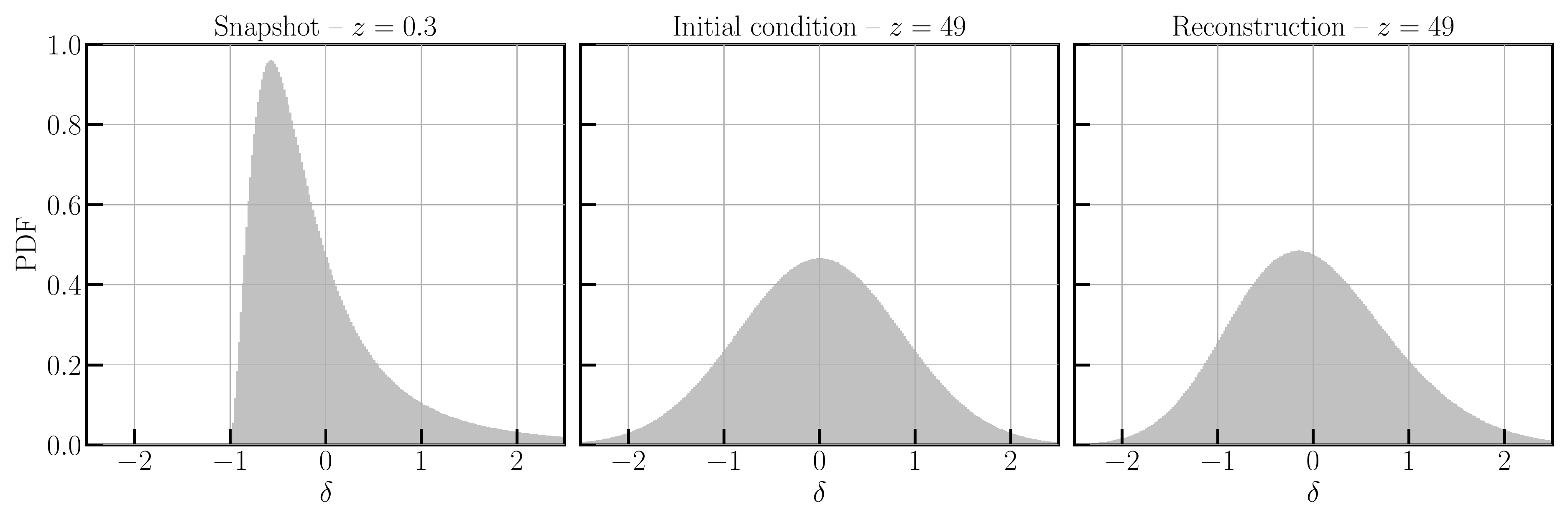}
\caption{Histograms of the density contrast $\delta$ corresponding to the images depicted in Fig.~\ref{fig:threedensity}.}
\label{fig:threehistogram}
\end{figure*}

As a second qualitative diagnostic we show the 1-point distribution function of the smoothed densities in all three, original, initial, and reconstruction, see Fig.~\ref{fig:threehistogram}. The left, $z_s=0.3$, panel highlights the known skewed distribution of the density contrast in the non-linearly evolved universe with highly collapsed overdensities among a mostly underdense matter distribution. This stands in contrast to the (by construction) Gaussian distribution seen at $z_f=49$. As is intuitive, our reconstruction leads to an almost Gaussian density contrast, that exhibits only weak skewness due to residuals of the very non-linear structures at the input reshift. The idea to morph the skewed density contrast to become more Gaussian has indeed inspired some of the first variants on reconstruction methods, so-called Gaussianization techniques~\citep{1992MNRAS.254..315W}.  Comparisons of 1-point distributions of true and reconstructed density contrasts can also be found in {\it e.g.} \citet{2017PhRvD..96b3505S}.

\subsection{Cosmological quantities}
\label{sub:cosmological quantities}

We here characterize the power spectra and correlation functions of our reconstructions by comparing them with their linear expectations, both individually and after averaging.  We account for various noise contributions, such as sample variances, shot noise, and conventional broad-band power contributions, we finally demonstrate the reconstructions' excellent agreement with their expectation.

\subsubsection*{Power spectra}

\begin{figure*}
    \centering
    \includegraphics[width=\columnwidth]{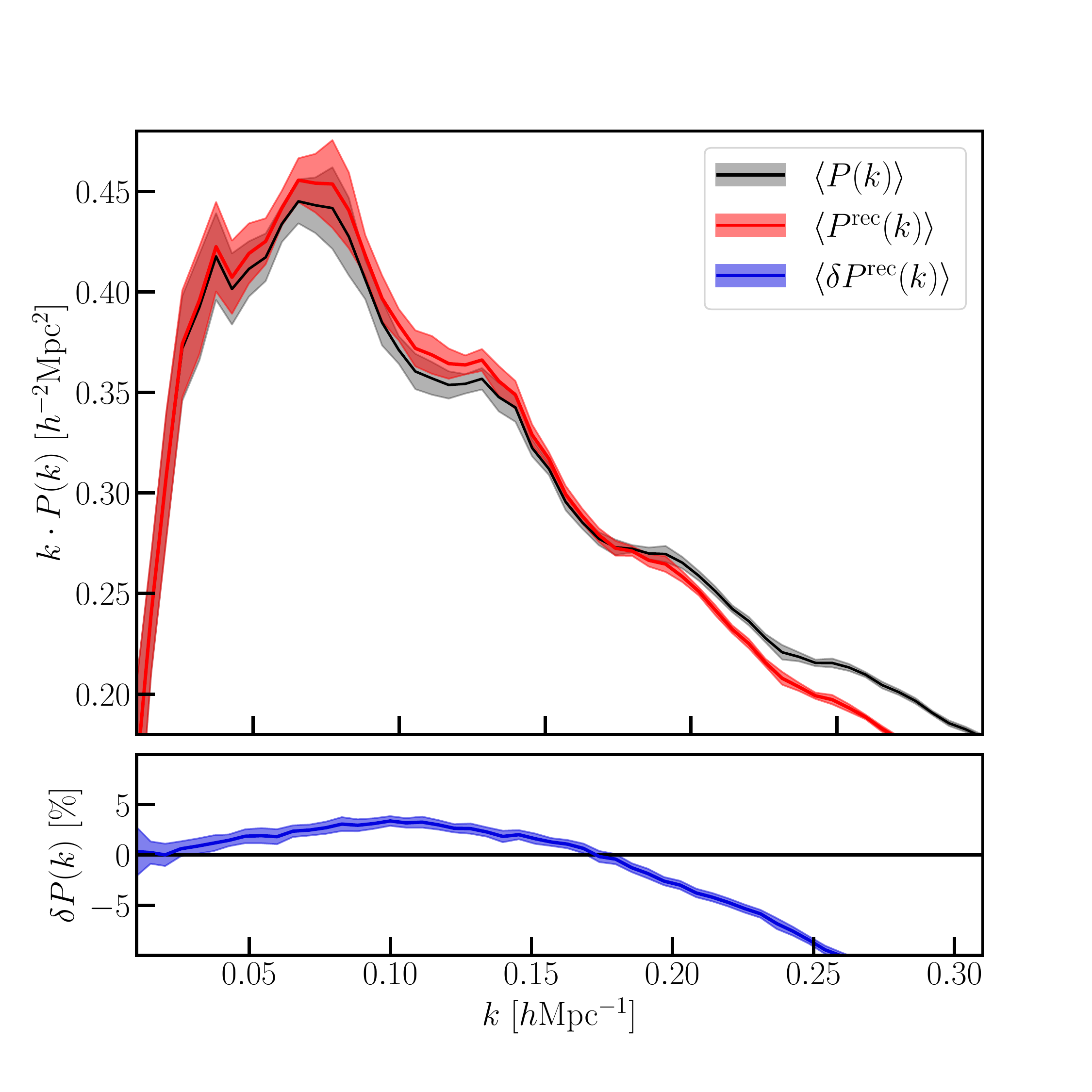}
    \includegraphics[width=\columnwidth]{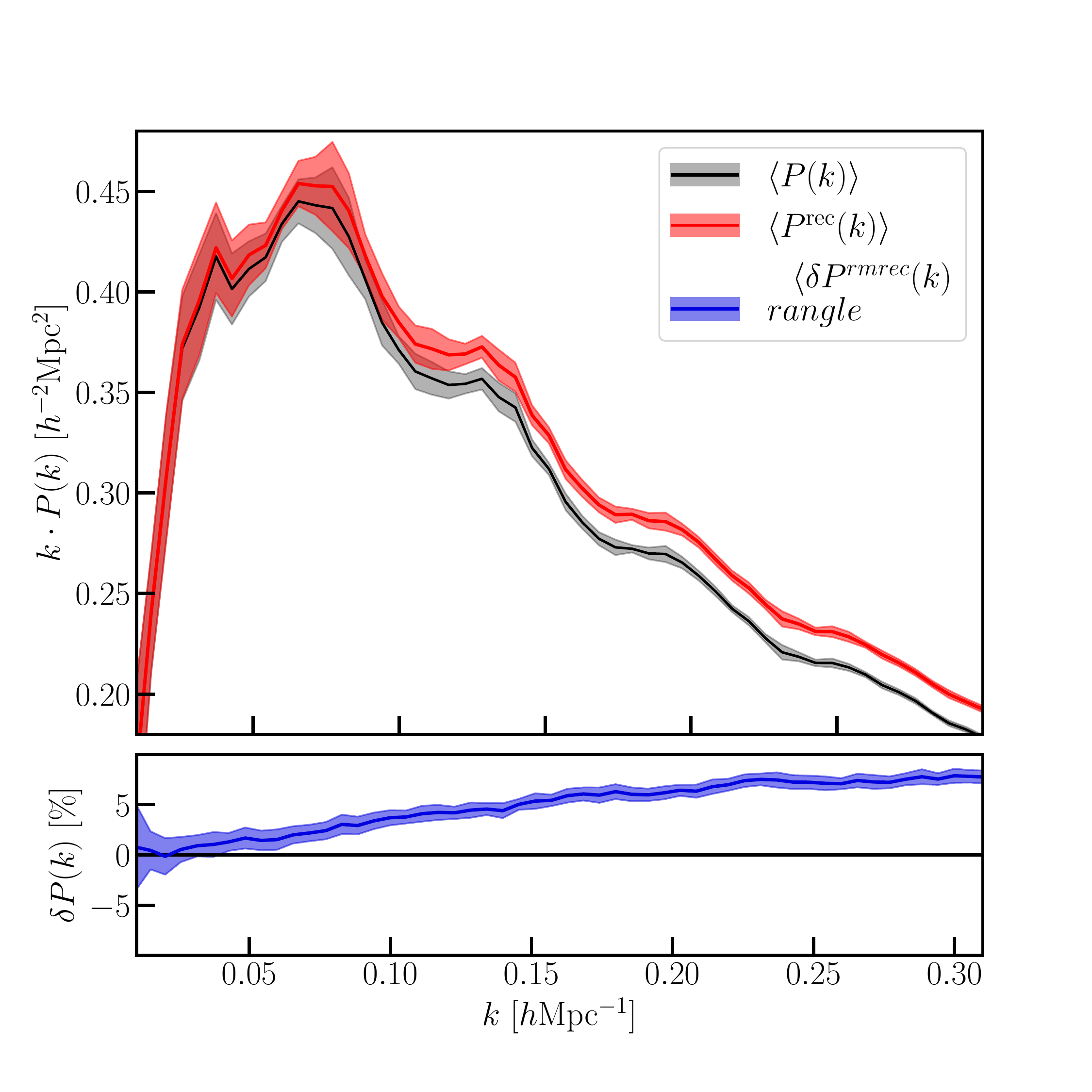}
    \caption{Average power spectra of ten \textsc{AbacusCosmos} simulations and their reconstructions at redshift $z=49$.  Reconstructions were performed beginning with samples at redshift $z_s=0.3$ ({\it left}) and $z_s=1.5$ ({\it right}).  {\it Upper panels:} Average power spectra and standard deviations for simulations and reconstructions, respectively. {\it Lower panels:} Averaged relative differences of reconstructed and true power spectra, cf. Eq.~\ref{eq:powerspectrumdeviation}.  Shaded bands show 1$\sigma$ deviation.}
    \label{fig:powerspectrum}
\end{figure*}

Even though the ultimate goal is to obtain a good estimate of the true power spectrum, that underlies any particle sample, `cosmic sample' variance -- arising from selecting a particular box of the simulated universe -- is an unyielding obstacle inherent to any such endeavor.  However, for the sake of inspecting only the power induced (or deduced) by our reconstruction algorithm, we here compare the power spectra of simulation and reconstruction for each simulation, respectively, thereby artificially circumventing cosmological sample variance.  This approach might prove helpful in future work, when biases potentially present in our reconstruction, should be removed from reconstructions of real data.  A second source of sample variance -- in the following referred-to as `subsample' variance -- appears when sampling the density field with a finite number of particles.  To avoid the false impression of not recovering the expected power on large scales, we here create a set of five subsamples per simulation, each with the same number of particles, yet selected with different random seeds.
For each simulation $i$ and subsample $j$ we compute the power spectrum $P^{\rm rec}_{ij}(k)$ of the sample reconstructed to redshift $z_f=49$ and compare it with the initial power spectrum $P_i(k)$ at the same redshift as follows.
\begin{equation}
    \delta P^{\rm rec}_{ij}(k) := \frac{P^{\rm rec}_{ij}(k)}{P_i(k)}-1
    \label{eq:powerspectrumdeviation}
\end{equation}
This quantity is then averaged over all $i$ and all $j$.  We show this average and its standard deviation in the bottom panels of Fig.~\ref{fig:powerspectrum} below the average power spectra for both initial conditions and reconstructions from $z_s=0.3$ (left panel) and $z_f=1.5$ (right panel), after having corrected for shot noise, as expanded in appendix~\ref{app:psdetails_shotnoise}.\footnote{Even though, as will be seen below, the influence of shot noise is usually removed by subtracting functions describing anomalous broad-band power~\citep{Seo:2008yx} we chose to isolate the computation of shot noise first, in order to disentangle noise that scales with particle number from other effects intrinsic to our method.}  In addition, and by the example of one simulation we show, in appendix~\ref{app:psdetails_subsamplevariance}, the effect of subsample variance.

Already at this stage, the averages and the standard errors of the reconstructed power spectra agree with the linear expectation at the $\sim5\%$-level at wavenumbers $k\lesssim0.2\,h\textrm{Mpc}^{-1}$ for both starting redshifts.  The relative deviations of reconstructed and initial power spectra seen in the bottom panels exhibit a smooth dependency on $k$ that indeed is known to arise from mode-coupling in the non-linear regime~\citep{1999MNRAS.308.1179M,1999ApJ...527....1S,Crocce:2007dt,Seo:2008yx,Xu:2012hg}.  Hence, and in line with other approaches for initial density reconstruction~\citep[\textit{e.g}][]{Xu:2012hg}, we introduce a broad-band term to our reconstructed power spectra, that, after fitting to the corresponding initial power spectrum, $P_i(k)$, compensates for this deviation:
\begin{equation}
    P_i(k) =  B(k) \cdot P_{ij}^{\rm rec}(k),
    \label{eq:broadbandnoise}
\end{equation}
where we found $B(k)=B_0 + B_1\cdot k + B_2\cdot k^2$ to be sufficient in accounting for the observed discrepancies when performing the fit in the range shown in the figures.  The data shows no further evidence for introducing additive corrections instead or in addition, and also gives no support for higher-order terms in $B(k)$.  This supports our understanding of shotnoise contributions as well as their sufficient subtraction as above.  While the linear growth factors in Eq.~\ref{eq:zeldovichapprox2} were chosen to match exactly those of the simulations, we expect $B_0$ to be close to 1.  Indeed we find this to be the case, and for completeness list the fitted values in the table below.

$$
\begin{array}{c|cc}
        &   z_i=0.3 &   z_i=1.5  \\ 
\hline        
    B_0 &    1.01   &   1.01    \\
    B_1 &   -0.86   &    -0.43  \\ 
    B_2 &   4.55    &   0.53    \\
\end{array}
$$

It should be noted that in practice the fitting is done by considering templates of linear power spectra that are each generated with a set of cosmological parameters, and are subsequently modified to include effects of non-linear growth.  Since the intention of this paper is simply to show the accuracy of recovering the expected power of linear fluctuations, instead of recovering cosmological parameters, we chose to fit directly to the power spectrum $P_i(k)$ of the $z_f=49$ density.

While the accounting of broad-band power is also necessitated by a combination of effects, such as redshift-space distortions and surveying effects, we must attribute the power discrepancy to the reconstruction itself.  In practice, however, such nuisance terms would absorb any such unexpected power, regardless of its source.

\begin{figure*}
    \centering
    \includegraphics[width=\columnwidth]{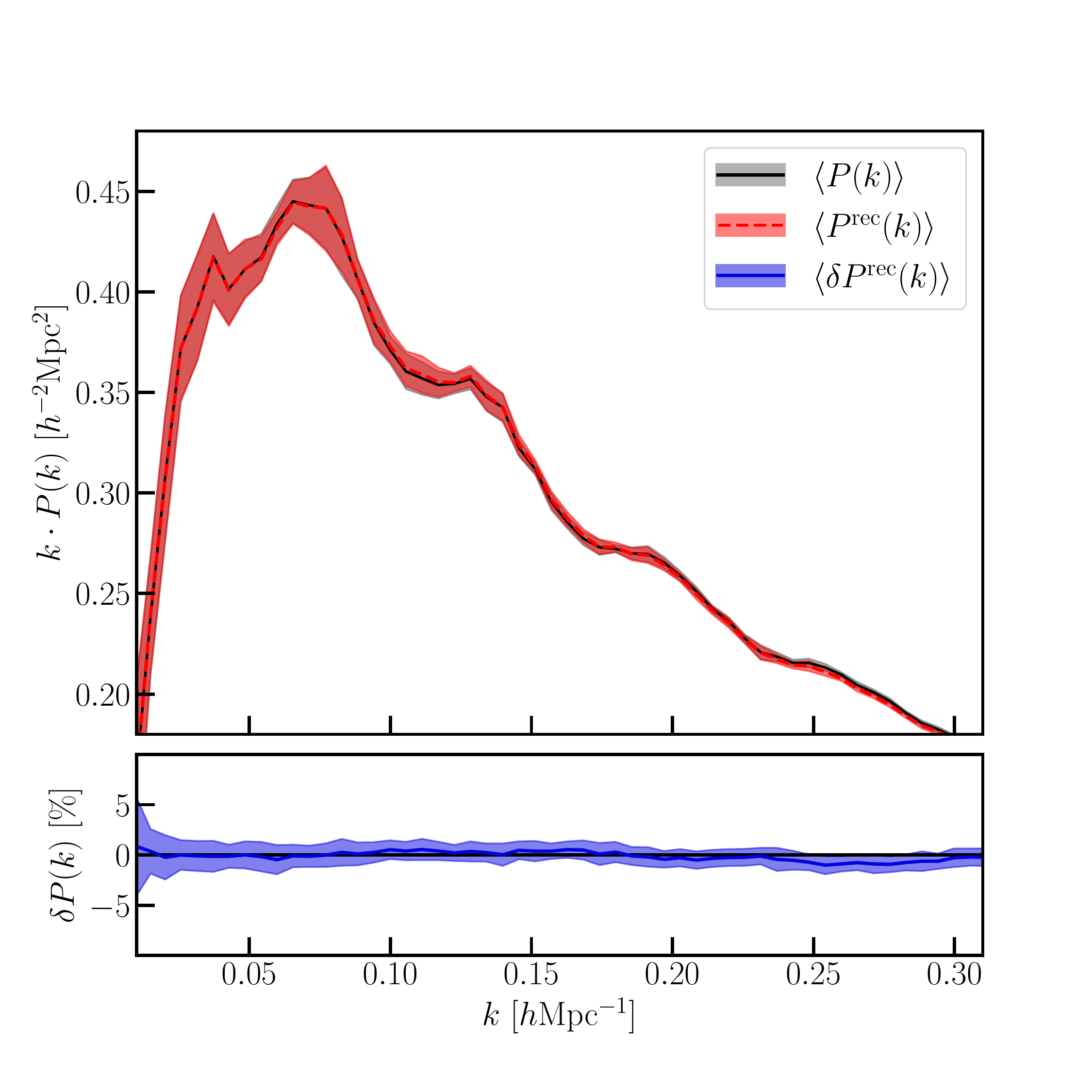}
    \includegraphics[width=\columnwidth]{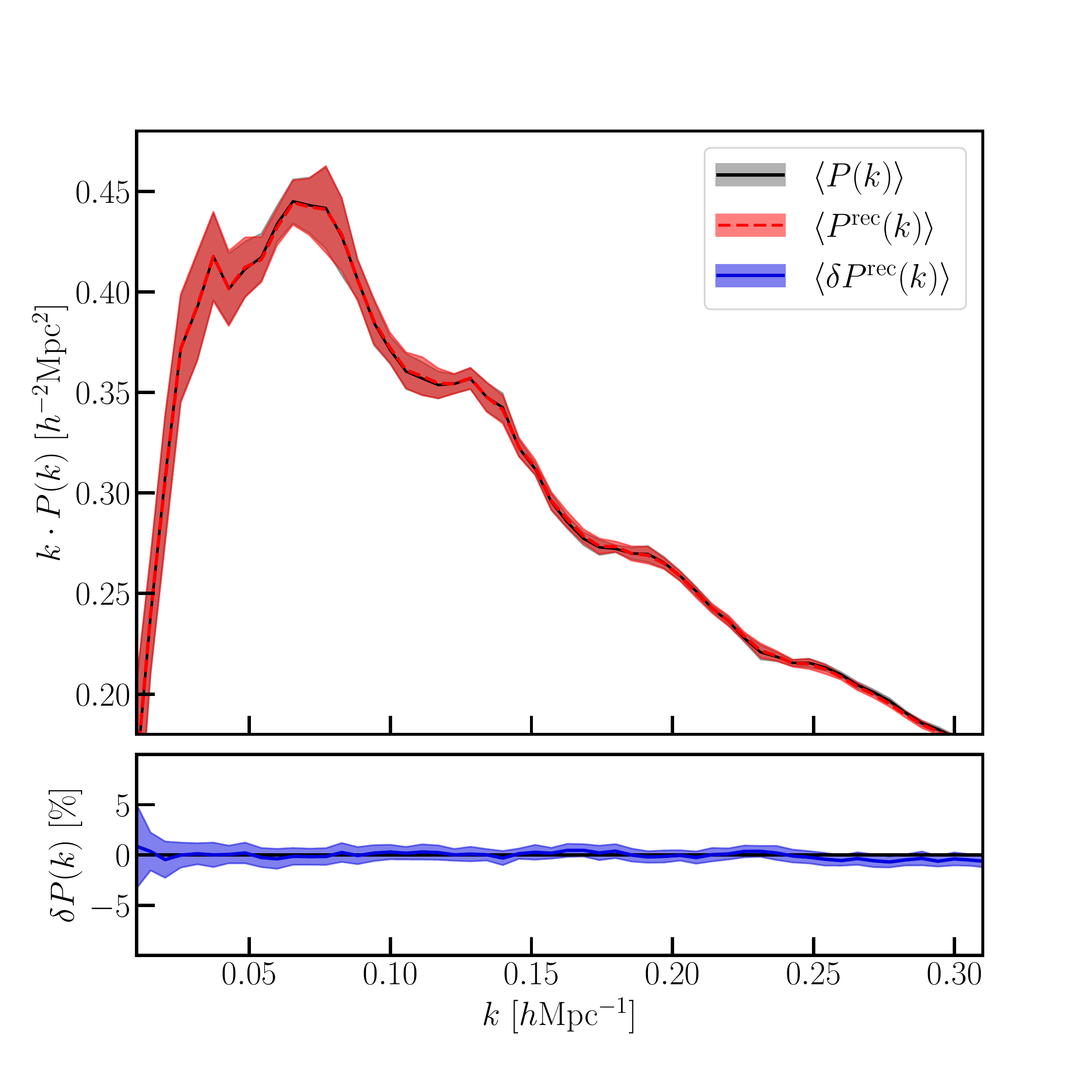}
    \caption{Same as Figure~\ref{fig:powerspectrum} but after accounting for broad-band noise, cf.~Eq.~\ref{eq:broadbandnoise}.}
    \label{fig:powerspectrum_fitted}
\end{figure*}

Figure~\ref{fig:powerspectrum_fitted} shows the relative differences after having included the broad-band terms.  The range of $k$ over which the reconstructed power matches that of the initial density is striking, reaching $k\approx 0.4\,h\rm{Mpc}^{-1}$ ($0.5\,h\rm{Mpc}^{-1}$) for $z_s=0.3$ ($z_s=1.5$) before deviating past the $5\%$-level.

\subsubsection*{Correlation functions}

\begin{figure*}
    \centering
    \includegraphics[width=\columnwidth]{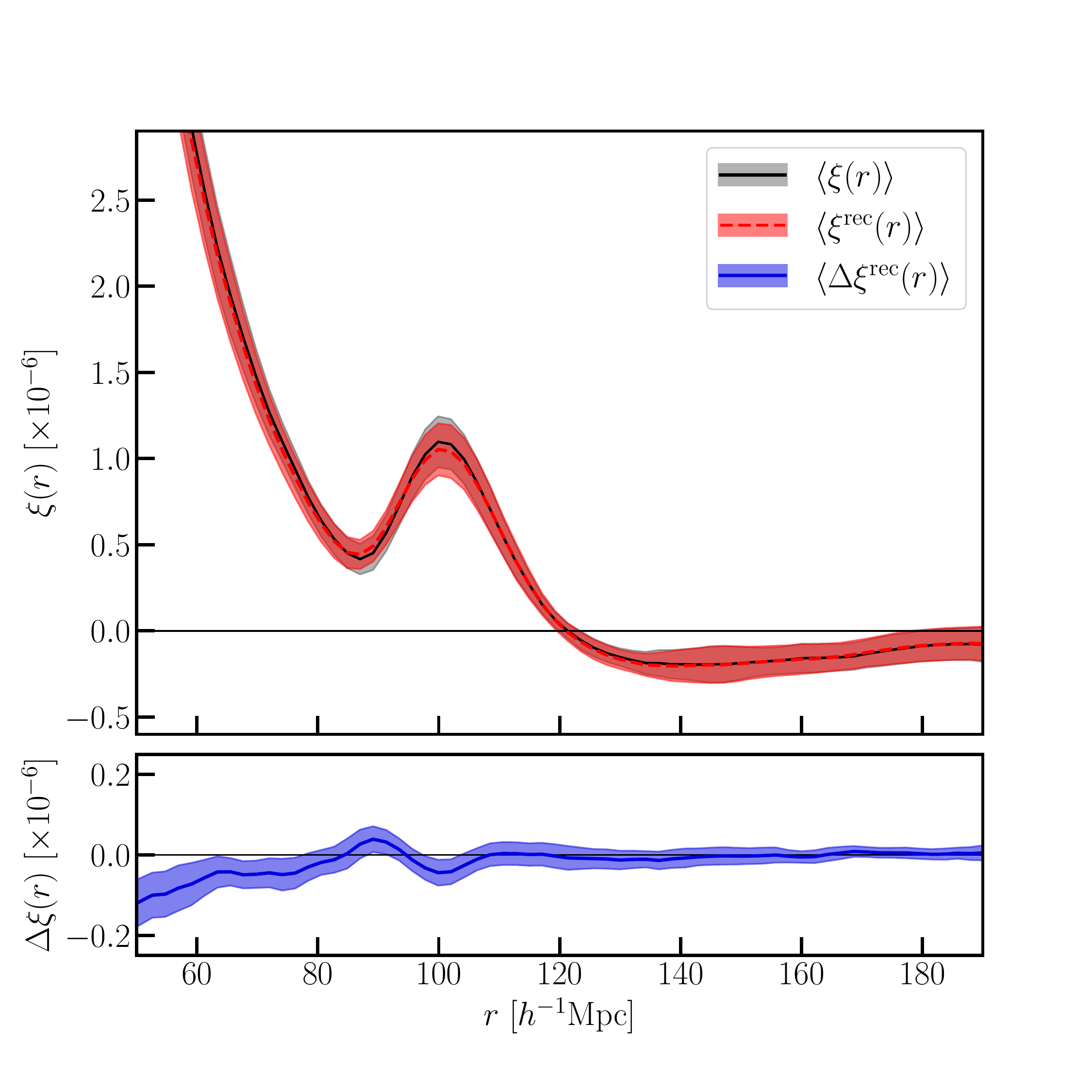}
    \includegraphics[width=\columnwidth]{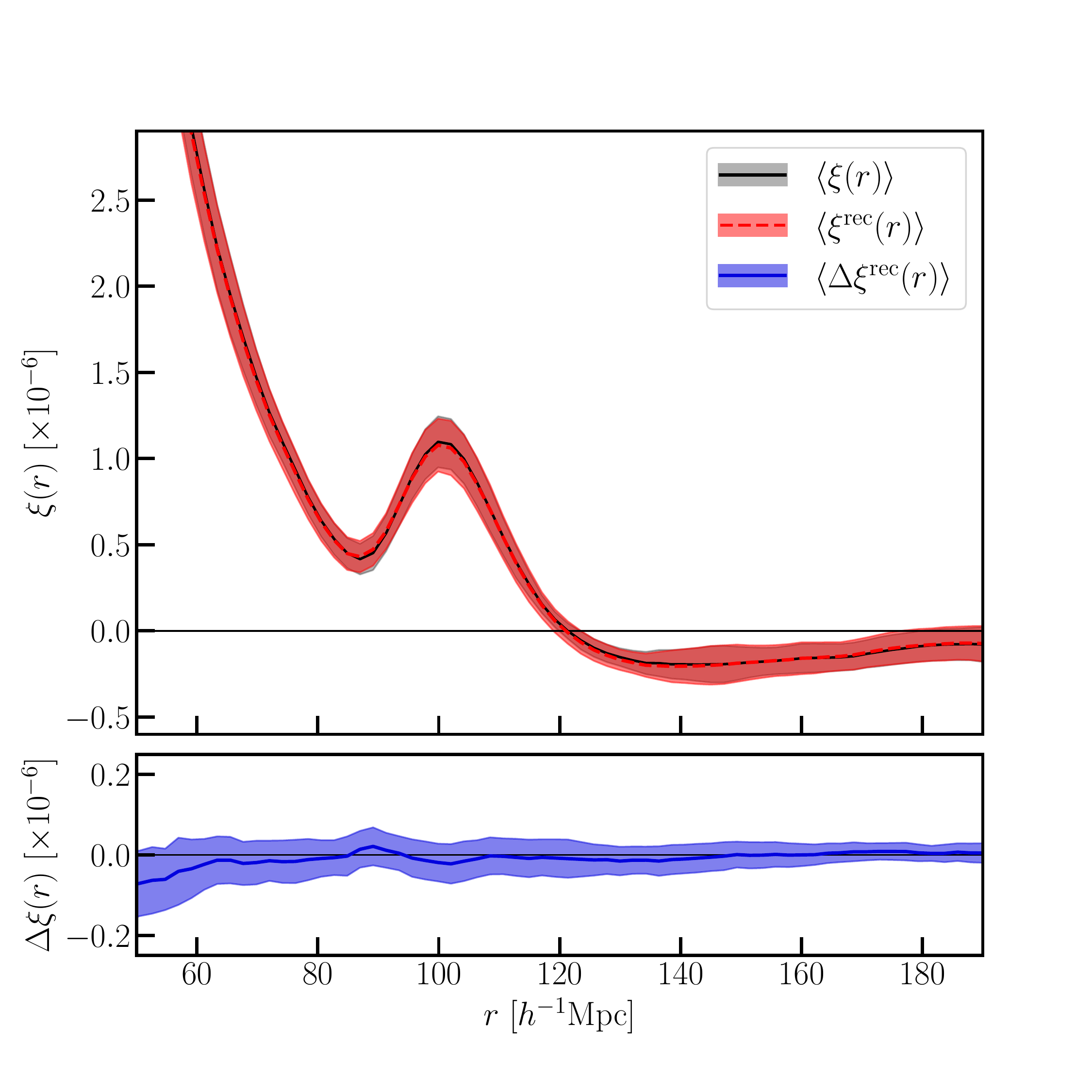}
    \caption{Average correlation functions of ten \textsc{AbacusCosmos} simulations and their reconstructions at redshift $z=49$.  Reconstructions were performed beginning with samples at redshift $z_s=0.3$ ({\it left}) and $z_s=1.5$ ({\it right}).  {\it Upper panels:} Average power spectra and standard deviations for simulations and reconstructions, respectively. {\it Lower panels:} Averaged differences of reconstructed and true correlation functions, cf. Eq.~\ref{eq:correlationfunctiondeviation}.  Shaded bands show 1$\sigma$ deviation.}
    \label{fig:correlationfunction}
\end{figure*}

Visually more revealing of the BAO signal is the two-point correlation function (the Hankel transform of the power spectrum).  We repeat previous exercise for the correlation functions computed from simulations and reconstructions.  Instead of the relative difference, as above, we here compute the absolute difference for each pair $ij$ of simulation and reconstruction.
\begin{equation}
    \Delta\xi_{ij}(r) := \xi^{\rm rec}_{ij}(r)-\xi_{i}(r)
    \label{eq:correlationfunctiondeviation}
\end{equation}
Correlation functions, their spread in simulations and reconstructions, and the absolute difference are shown in Fig.~\ref{fig:correlationfunction}.  The BAO peak and its shape are recovered well and with uncertainties comparable to the sample variance of the initial conditions themselves.  As before, a discrepancy growing towards low separations $r$ is removed by fitting for broad-band influences in the shown range. As opposed to broad-band noise in the power spectrum, we here find no evidence for any scale-dependencies and simply allow for a constant bias, again only performing the fit in the shown range,
\begin{equation}
    \xi_i(r) = B\cdot\xi^{\rm rec}_{ij}(r).
    \label{eq:broadbandnoise_cf}
\end{equation}

$$
  \begin{array}{c|cc}
        &  z_i=0.3 & z_i=1.5 \\
    \hline    
    B   &  1.025   & 1.011    
  \end{array}
$$
 
\begin{figure*}
    \centering
    \includegraphics[width=\columnwidth]{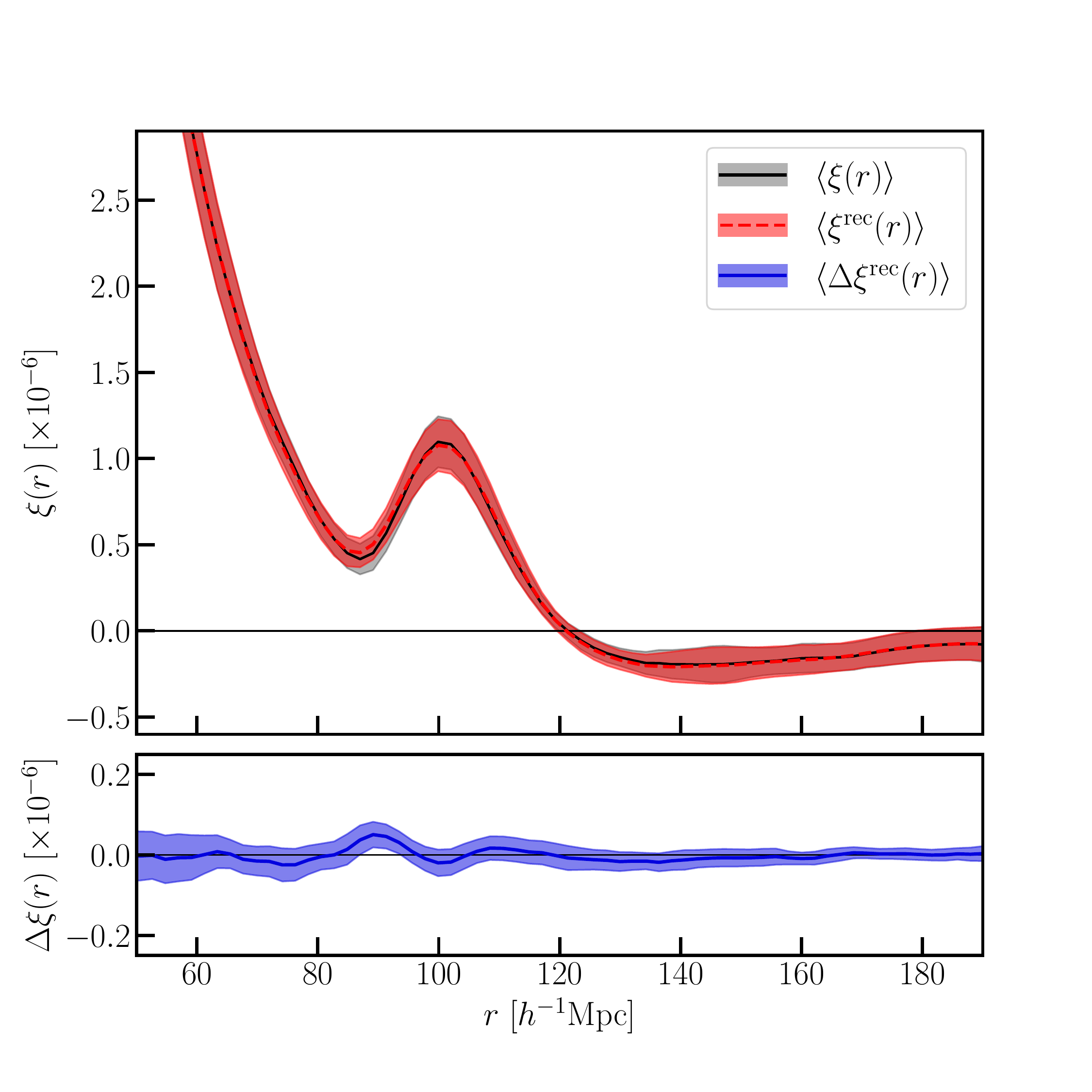}
    \includegraphics[width=\columnwidth]{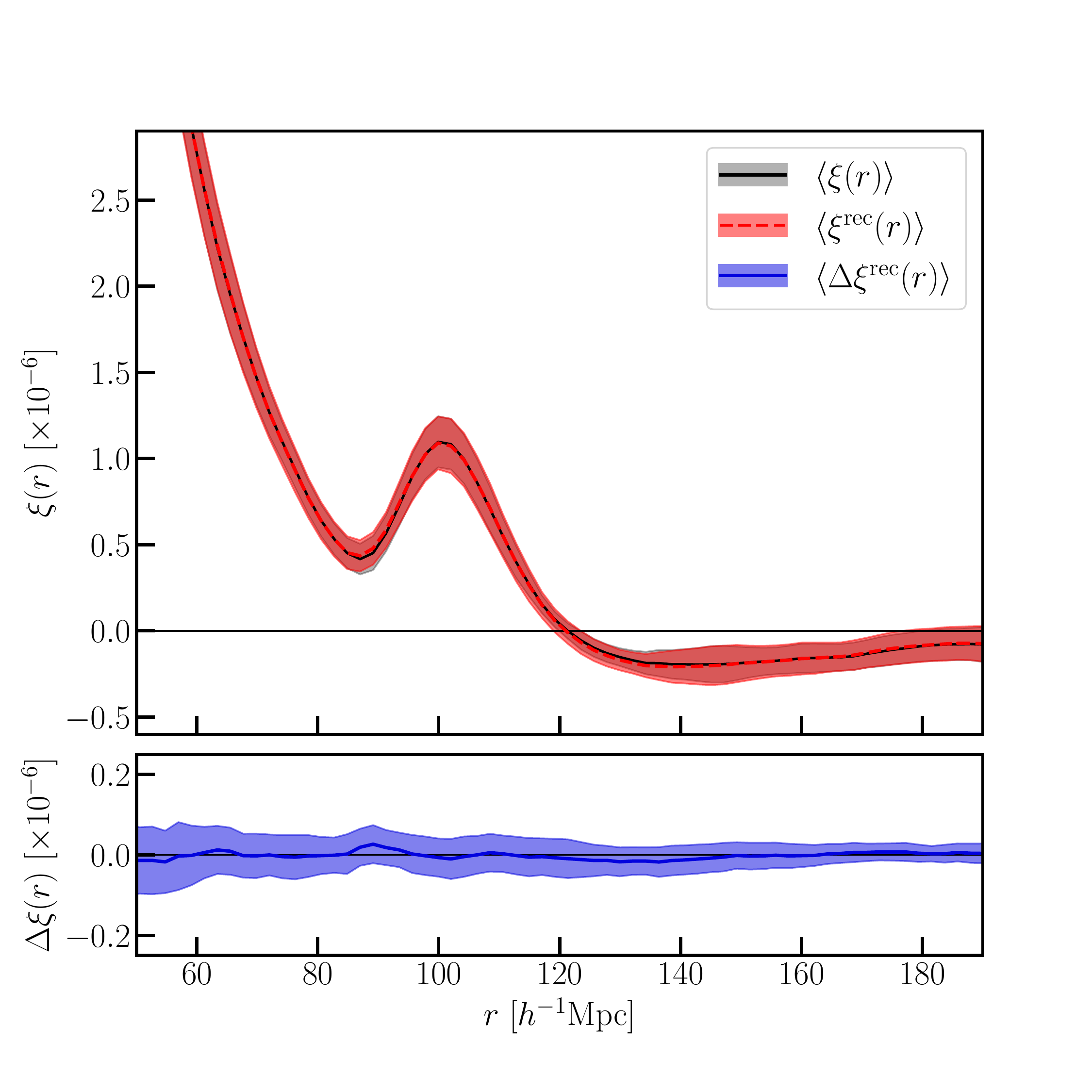}
    \caption{Same as Figure~\ref{fig:correlationfunction}, but after accounting for broad-band noise, cf.~Eq.~\ref{eq:broadbandnoise_cf}.}
    \label{fig:correlationfunction_fitted}
\end{figure*}

The corresponding average correlation functions are shown in Fig.~\ref{fig:correlationfunction_fitted}.  A slight ($\sim\!1\sigma$) deviation around the BAO feature reveals residual dampening of the peak for the $z_s=0.3$ reconstructions, which, however, is not significant for those reconstructions beginning at the higher redshift, $z_s=1.5$.

\begin{figure*}
    \centering
    \includegraphics[width=\columnwidth]{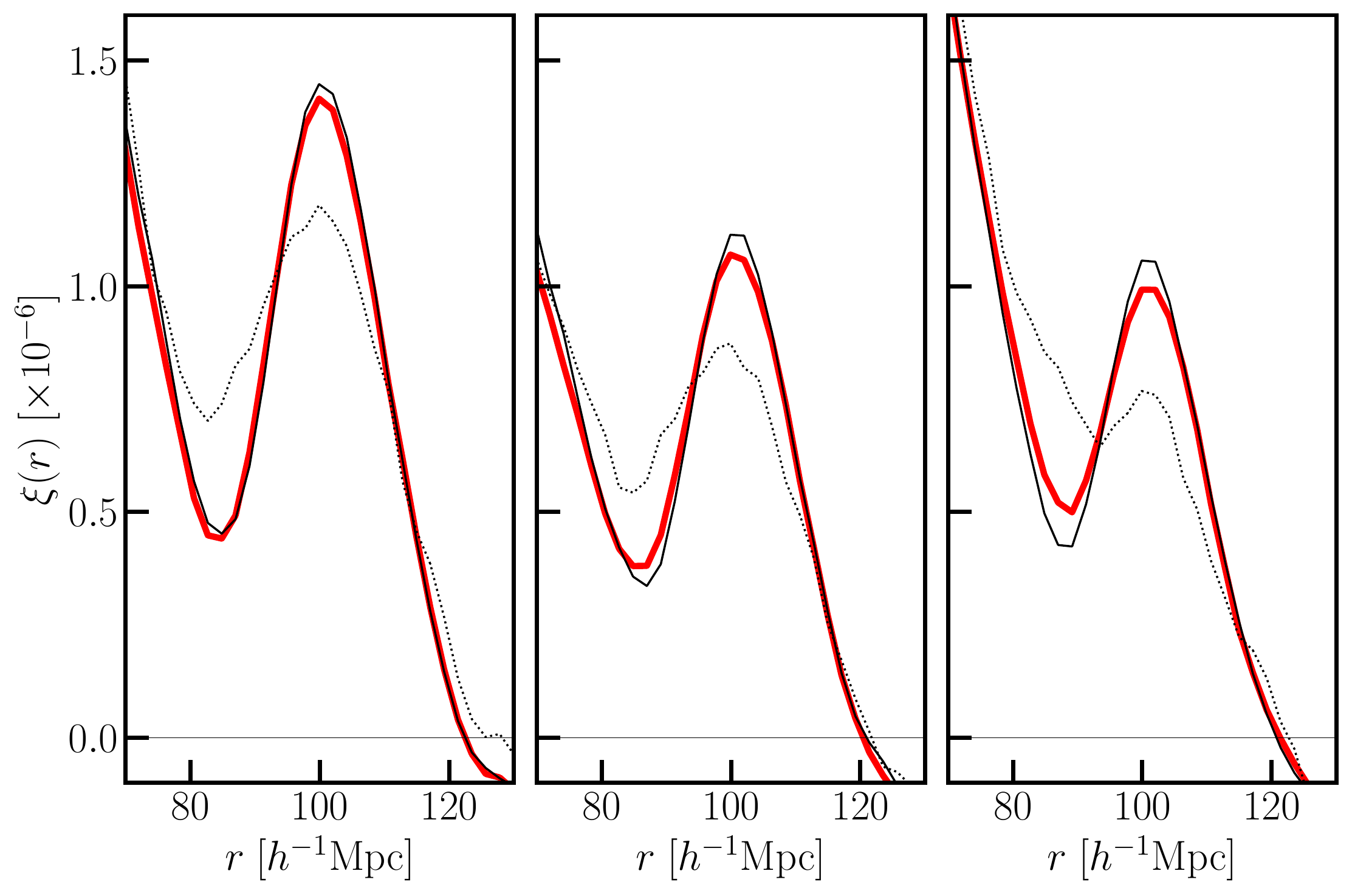}
    \includegraphics[width=\columnwidth]{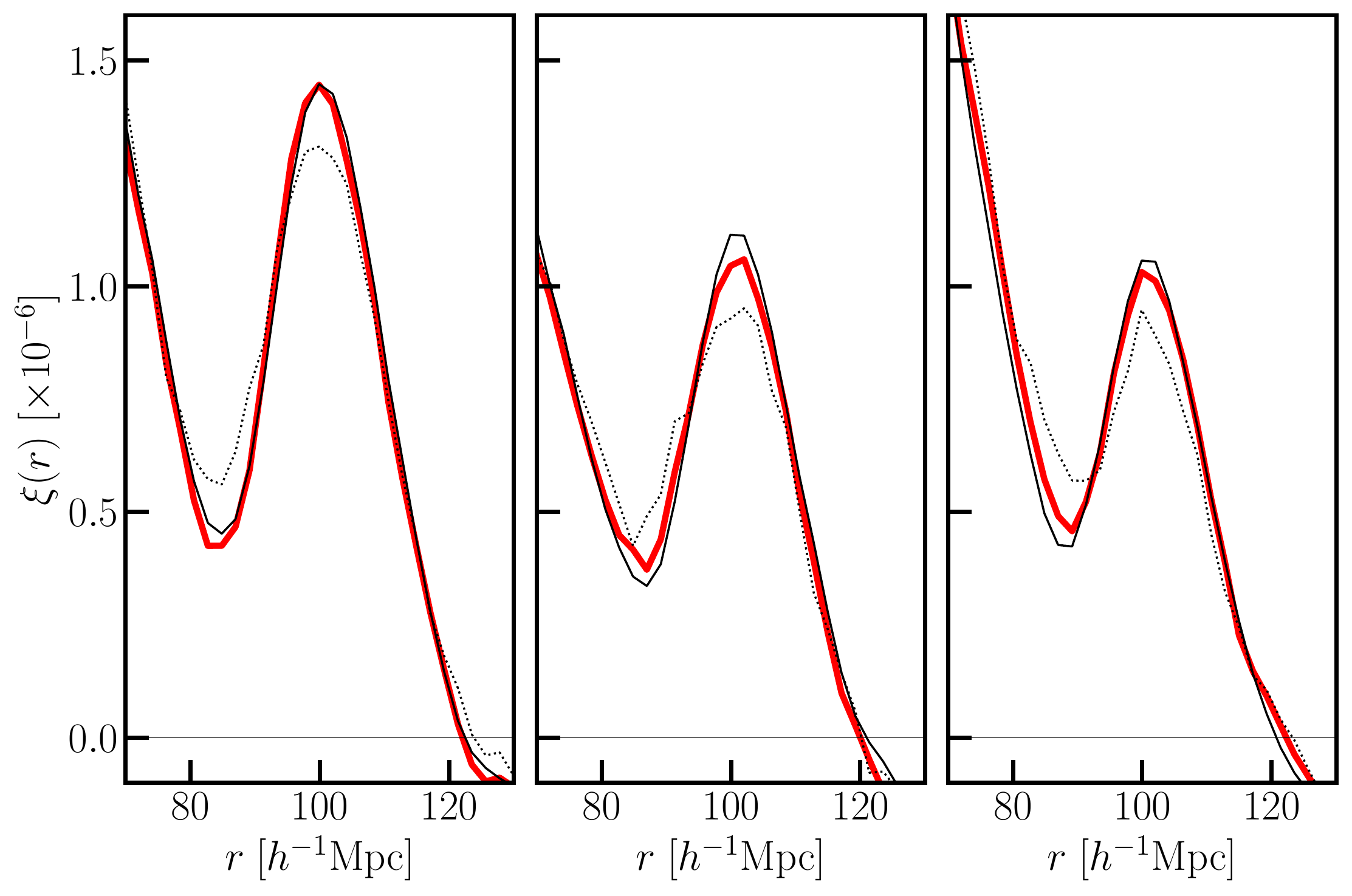}
    \caption{Examples of correlation functions in three different simulations and their reconstructions.  The correlation functions as measured in snapshots at $z_s=0.3$ ({\it dotted black}) and $z_f=49$ ({\it solid black}) are compared with the reconstruction at $z_f=49$ ({\it red}).  All curves have been scaled to match the linear growth amplitude at $z=49$ for visual comparison.}
    \label{fig:3correlationfunction}
\end{figure*}

To demonstrate the ability of our reconstruction algorithm on individual cases of correlation functions, we show in Fig.~\ref{fig:3correlationfunction} three different examples, in which we compare the correlation functions of the reconstructed densities with those of the initial densities, as well as those of the samples with which the reconstruction algorithm was fed.  Due to cosmic sample variance in the simulated volume, the BAO feature visible at $z_s=0.3$ or $z_s=1.5$ ({\it dotted black}) may be more or less pronounced ({\it left to right panels}).  However, in all cases, the reconstruction is able to recover well the precise shape of the BAO bump at the redshift of reconstruction, $z_f=49$, as shown by the close agreement between simulation ({\it solid black}) and reconstruction ({\it red}).  In all cases one can observe the well-known sharpening of the peak~\citep{Eisenstein:2006nj} that reconstruction methods generally aim for.  Especially the third example of each panel exhibits this effect, wherein the peak is hardly recognizable at $z_s=0.3$.

\subsubsection*{Accuracy of acoustic scale recovery}

Finally, we provide a tentative quantification of the accuracy with which the sound horizon at decoupling, {\it i.e.}~the position of the BAO peak, is recovered using the same simulations as above.  To this effect, we localize the peak of the correlation function in the vicinity of its input value, $r_s(z_{\rm drag})=99.09\,h^{-1}\rm{Mpc}$, after having interpolated the function with a cubic spline to smooth out discreteness effects.  This is done for both the initial correlation functions $\xi_i(r)$, leading to $r^{\rm IC}_s$, as well as the reconstructed correlation functions $\xi^{\rm rec}_{ij}(r)$ from both input redshifts, $z_s=0.3$ and $z_s=0.5$, resulting in $r^{\rm rec}_s$.  This results in 50 peak positions each.  For both redshifts we then compute the mean and standard deviation of the fractional deviation
\begin{equation}
   \delta r_s = \frac{ r^{\rm rec}_{s,ij}-r^{\rm IC}_{s,i} }{ r^{\rm IC}_{s,i}},
\end{equation}

$$
  \begin{array}{c|cc}
        &  z_i=0.3 & z_i=1.5 \\
    \hline  
   \overline{ \delta r_s}\pm \sigma \;[\%]  & -0.04 \pm 0.37    & 0.08\pm0.52
  \end{array}
$$

Even with this rough attempt, we are able to recover the BAO scale with sub-percent scatter.  We will return to this point in future, dedicated works containing more comprehensive methods for the localization of the BAO peak that take into account the full shape of the peak, cf.~\citet{2017PhRvD..96b3505S}, and including comparisons with existing reconstruction algorithms.

\section{Conclusions}
\label{sec:conclusion}

To extract informations on the early Universe from the present Universe requires the undoing of the nonlinear growth of structures. This complex inverse problem, sometimes referred to as the cosmological reconstruction problem, is often tackled in a forward iterative manner: essentially an initial model with a set of parameters is assumed and simulated and then compared to the data at the present epoch within an iterative loop till the best match between the model and the data is achieved \emph{statistically}. In previous works we have shown that the cosmological reconstruction problem is a subset of the general class of mass transportation problems that are solved through optimal transport theory and as such can be tackled deterministically to yield a unique solution. The problem is well-formulated by the Monge-Ampère equation.
Our previous solution to the Monge-Ampère equation was obtained through a \emph{combinatorial} fully-discrete algorithm, which explored a huge solution space to find the optimal assignment.  These algorithms have a complexity of $ N^3$, for a dataset with $ N$ points, which renders them impractical for applications to big data in cosmology. Although there has been reports of faster variants of combinatoric algorithms  with a complexity of at best $N^2\log ( N)$ (see review in \cite{OTReviewMerigotThibert}), their performance however remains too slow for large-scale cosmological problems.

In this article, we have presented a new \emph{semi-discrete} algorithm which makes a direct use of the \emph{variational} nature of the cosmological reconstruction problem. It finds a quick path to the solution by fully exploiting the first and second order derivatives of the objective function (that is both \emph{smooth} ($C^2$) and \emph{convex}). This is made possible by a fortuitous yet elegant convergence between the physical, mathematical and computational aspects of the problem: the specific cosmological setting that we considered (continuous mass transported to a pointset) has nice mathematical properties--
Monge-Ampère equation translated into a smooth and convex optimization problem, with an underlying geometric structure (Laguerre diagram) that can be \emph{exactly} computed by our algorithm. Our semi-discrete algorithm has a complexity of $N \log(N)$ which makes it significantly more efficient than any combinatoric algorithm.

As a concrete example of a practical consequence, our previous combinatoric code could
reconstruct a typical $128^3$ $\Lambda$CDM dark-matter-only simulation in a month on a desktop computer station. In contrast, with our new semi-discrete algorithm described in this article, the same reconstruction can be done on a portable personal computer in less than five minutes. 

Subtle informations in the primordial power spectrum, such as BAO, can only be retrieved through the analyses carried on horizon scales and on hundres of millions of particles: a task completely unfeasible through combinatorial algorithms. Our new semi-discrete algorithm processes such massive datasets within hours, which makes it a powerful tool to reconstruct BAO and consequently test the theory of general relativity at cosmological scales. Here, we have chosen BAO measurement as a challenge to test the power of our algorithm, however it is needless to say that other decisive features of the primordial power spectrum, {\it e.g.} the primordial non-Gaussianity can also be detected by our algorithm and consequently we can also provide constraints for the inflationary models \citep{2006MNRAS.365..939M}. 

Aside from the BAO and primordial non-Gaussianity, our deterministic algorithm can also recover the velocity field, including the relevant phase informations. Hence it can provide priors for Bayesian predictions of large-scale structure. The density and velocity field information in the reconstructed initial conditions can be used to develop Bayesian priors to probe CMB polarization and temperature anisotropy maps for evidence of any phase anomalies. This potentially provides a powerful new insight into the validity of the standard cosmological model. 

Our reconstruction method holds for as long as the convexity holds which implies that the reconstructed map between initial and final distribution remains valid into the nonliner regime and at least up to the third-order in the Lagrangian perturbation theory. However, in extracting the density field from these maps, we have used the Zel'dovich approximation for convenience as it gives us a simple analytic expression (\ref{eq:zeldovichapprox2}). The reason why higher-order terms have not been implemented here is that for BAO reconstruction a broad-band fitting function is often used to account for mode-coupling as well as other effects. We have shown that our BAO reconstruction enjoys a high sub-percent accuracy with the least number of fitting polynomial parameters. In the forthcoming work we shall implement higher orders which could reduce the need to fitting parameters and rendre the BAO reconstruction model-independent.

In this work, our code ran on particle samples where all points have the same mass and reside in real space within periodic boundary conditions.
In the forthcoming works, we shall generalize this computational setting to other geometrical configurations, more relevant for observational surveys. We shall account for nonlinear halo bias, adapt our code to redshift space and to more general boundary conditions and survey geometries and make it available for applications to data expected from the future telescopes.


\section*{Acknowledgements}

The authors thank Jean-Michel Alimi, Jean-David Benamou, Enzo Branchini, Yann Brenier, Quentin Mérigot, Joe Silk for discussions, and Alain Filbois who developed an early prototype of the parallel triangulation code. SvH thanks Lehman Garrison for assistance with the AbacusCosmos simulation suite. This work was supported by the EXPLORAGRAM Inria AeX grant.  For analyzing the results, this work made use of the \texttt{numpy}, \texttt{matplotlib}, \texttt{nbodykit}~\citep{nick_hand_2018_1336774}, and \texttt{hmf} python packages.

\newpage


\bibliographystyle{mnras}
\bibliography{SDMAK}


\appendix
\section{Algorithmic details}

\label{sec:algo}

\subsection{Numerical aspects}


Newton's algorithm for semi-discrete optimal transport can be summarized as follows:
$$
\begin{array}{ll}
   \mbox{\bf Input:}    &- \mbox{ the set of } N \mbox{ points } \bx_i \in V = [0,1]^3 \\
                        &- \mbox{ the masses } \mu_i \mbox{ such that } \sum_i \mu_i = 1 \\
                        &  \quad \mbox{ ( for instance, } \mu_i = 1/N  \mbox{)}\\[2mm]
   \mbox{\bf Output:} &- \mbox{the (unique) vector } \psi \in \mathbb{R}^N \mbox{ that maximizes } K(.) \\
                      &- \mbox{the Laguerre cells } (V_i^\psi)_{i=1}^N
                      \mbox{ that give for each } \bx_i \\
                      & \mbox{ the (continuous) set of } \bq \mbox{ points mapped to } \bx_i.

   \\[5mm]
                        \hline \\
   \ (1): & \psi \leftarrow [ 0 \ldots 0 ] \\
   \ (2): & \mbox{Loop} \\
   \ (3): & \quad \mbox{Compute the Laguerre diagram } (V_i^\psi)_{i=1}^N \\
   \ (4): & \quad \mbox{Compute the gradient } \nabla K(\psi) \\
   \ (5): & \quad \mbox{If } \| \nabla K(\psi) \|_\infty < \epsilon \mbox{ then Exit loop} \\
   \ (6): & \quad \mbox{Compute the Hessian matrix } \nabla^2 K (\psi) \\
   \ (7): & \quad \mbox{Solve for } \bp \in {\mathbb R}^n \mbox{ in  } \nabla^2 K(\psi) \bp = -\nabla K(\psi) \\
   \ (8): & \quad \mbox{Find the descent parameter } \alpha \\
   \ (9): & \quad \psi \leftarrow \psi + \alpha \bp \\
   \ (10): & \mbox{End loop}
\end{array}
$$

Figure \ref{fig:Newton} page \pageref{fig:Newton} shows some iterations of the Newton algorithm: the vector $\psi$ is optimized until each Laguerre cell $V_i^\psi$ has the prescribed area $\mu_i = 1/N$. \\

The algorithm above needs to compute multiple evaluations of the gradient and Hessian matrix of $K(.)$. The coefficients of the gradient and Hessian matrix can be deduced from the Laguerre
diagram $(V_i^\psi)_{i=1}^N$ that is computed at step (3). The associated algorithm is detailed later in the next subsection on the geometric aspects. 
Once the Laguerre diagram is computed, the  
coefficients ${\partial K}/{\partial \psi_i}$ of the gradient $\nabla K(.)$ are given by the following expression (\cite{DBLP:journals/corr/KitagawaMT16,DBLP:journals/cg/LevyS18}):
\be
  \frac{\partial K}{\partial \psi_i}  =  \mu_i - \int_{V_i^\psi} \rho\init(\bq) d^3\bq. 
\ee
In other words, this correspond to the mass $\mu_i$ associated with a point $\bx_i$ at present time, minus the mass of the matter that was transported there 
through the assignment map $\bx\now(.)$ (remember that the region transported to $\bx_i$
corresponds to $V_i^\psi$). For the vector $\psi$ that maximizes $K(.)$, all components of the gradient vanish, which means that each Laguerre cell $V^\psi_i$ has exactly the prescribed mass $\mu_i$. Since $\rho\init(.)$ is uniform, the integrated density over $V_i^\psi$ simply corresponds to the volume of $V_i^\psi$ (but the formula above is valid for an arbitrary $\rho\init(.)$ density).

This expression of the gradient leads also to a natural stopping criterion (line 5), the largest component of the gradient corresponds to the maximum error of transported mass. We stop the algorithm as soon as it is smaller than a prescribed $\epsilon$ (typically one percent of $\mu_i$, that is,  $\epsilon = 0.01/N$). \\

We now consider the Hessian matrix computed at step (6). Still following \citep{DBLP:journals/corr/KitagawaMT16,DBLP:journals/cg/LevyS18}, its coefficients are given by:
\ba
  \nonumber
  \frac{\partial^2 K}{\partial \psi_i \partial \psi_j} & = & \frac{1}{| \bx_j - \bx_i |} \int_{V_{ij}^\psi} \rho\init(\bq) d^3\bq \quad \mbox{if } i \neq j \\
  \frac{\partial^2 K}{\partial \psi_i^2} & = & - \sum_{j \neq i} \frac{\partial^2 K}{\partial \psi_i \partial \psi_j}
\ea
where $V_{ij}^\psi$ denotes the polygonal facet that is common to the Laguerre cell $V_i$ and $V_j$. Note that the Hessian matrix is sparse, and has a non-zero entry at coefficient $(i,j)$ if and only if the Laguerre cells $V^\psi_i$ and $V^\psi_j$ touch each other along a common facet. 

It can be remarked that the Hessian matrix coincides with the Finite Element $\mathbb{P}_1$ Laplacian. It can be explained as follows: the MA equation can be considered as a non-linear generalization of the Poisson equation, and its second-order expansion in the Newton algorithm naturally corresponds to the Laplacian. \\

\begin{figure}
    \centerline{
	   \includegraphics[width=0.5\columnwidth]{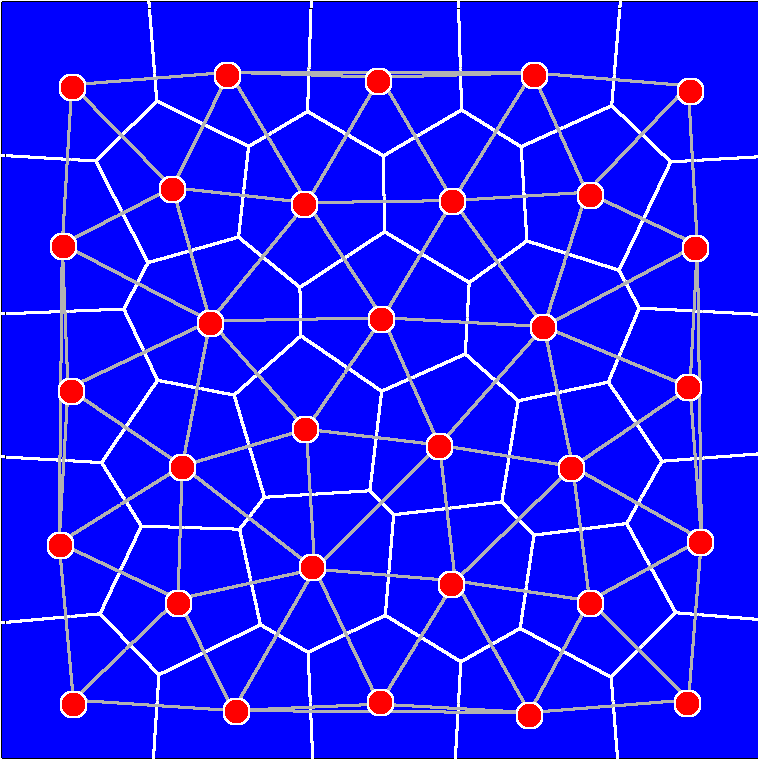}
    }
    \caption{Internaly, the Laguerre diagram is represented by its dual structure, called the regular triangulation (in gray).}
    \label{fig:triangulation}
\end{figure}

Step (7) of the algorithm computes the Newton step vector $\bp$, by solving a linear system. This linear system is typical of a Poisson equation discretized with Finite Elements, and can be solved using classical methods: we use the Conjugate Gradient algorithm \citep{Hestenes1952} pre-conditioned by Jacobi. The (sparse) Hessian matrix is stored using the Compressed Row Storage format, constructed and assembled using a specialized and highly optimized algorithm, interfaced with the Laguerre diagram. To tune the stopping criterion of the Conjugate Gradient algorithm, we used as a "ground truth" a direct solver (SuperLU) on small pointsets (thousands of points). For solving the linear system
$H \bp = -\bg$ we found that the stopping criterion $| H \bp - \bg | / | \bg | \le 10^{-3}$ results in nearly the same step vector $\bp$ as with the direct solver. 

We implemented two versions of the linear solver, a multi-threaded CPU version and a GPU version. On a high-end GPU (NVidia V100), the algorithm is typically 45 times faster (90 GFlops) than the multi-threaded CPU version (2 GFlops). \\

Once the step vector $\bp$ is computed, we need to find a good descent parameter $\alpha$. The KMT algorithm \citep{DBLP:journals/corr/KitagawaMT16}, provably convergent, works as follows:
$$
\begin{array}{ll}
   \ (1): & \alpha \leftarrow 1 \\
   \ (2): & \mbox{Loop} \\
   \ (3): & \quad \mbox{If  } \inf_i | V^{\psi + \alpha \bp}_i | >a_0 \\
   \ (4): & \quad \mbox{and } | \nabla K(\psi + \alpha \bp) | \le (1 - \alpha / 2) | \nabla K(\psi) | \\
   \ (5): & \quad \quad \mbox{then Exit loop} \\
   \ (6): & \quad \alpha \leftarrow \alpha / 2 \\
   \ (7): & \quad \mbox{Compute Laguerre diagram } (V^{\psi+\alpha \bp}_i)_{i=1}^N \\
   \ (8): & \mbox{End loop}
\end{array}
$$
where $a_0 = \frac{1}{2}\min\left( \inf_i  |V^0_i|  , \inf_i(\mu_i) \right)$ and where $|V_i^\psi|$ denotes the volume of a Laguerre cell. \\

The KMT algorithm iteratively halves the descent parameter $\alpha$ until two criteria are met: the volume of the smallest Laguerre cell needs to be larger than a threshold $a_0$ (line 3), and the norm of the gradient needs to decrease sufficiently (line 4). The threshold $a_0$ for the minimum Laguerre cell volume corresponds to (half) the minimum Laguerre cell volume for $\psi = 0$ (also called Voronoi diagram) and minimum prescribed area $\mu_i$ (in our case $1/N$). \\

Equipped with the KMT algorithm above, we can now compute the descent parameter $\alpha$, by plugging the algorithm above into line (8) of the Newton algorithm at the beginning of this section. \\

\begin{figure}
    \centerline{
	   \includegraphics[width=0.5\columnwidth]{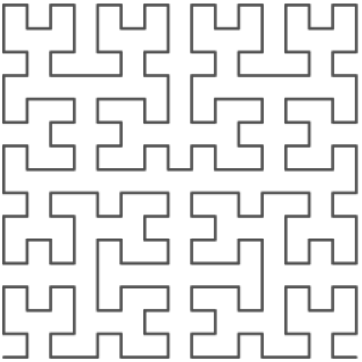}
	   \includegraphics[width=0.5\columnwidth]{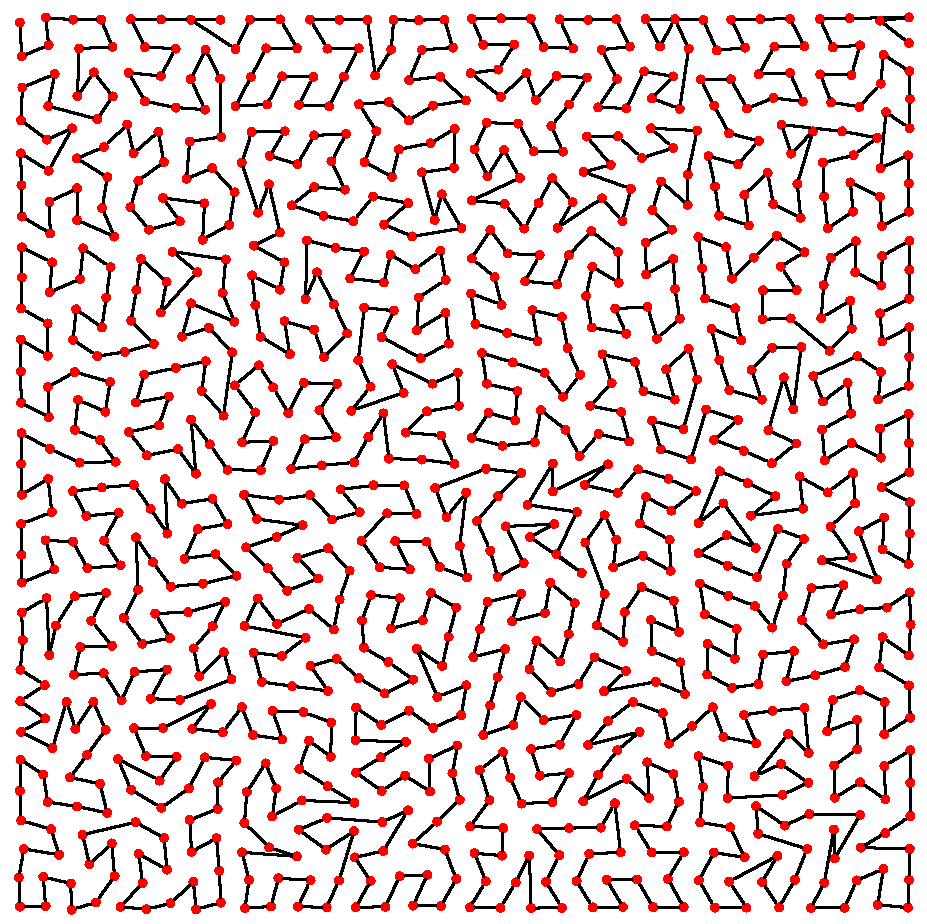}
    }
    \centerline{
       A \hspace{0.5\columnwidth} B 
    }
    \centerline{
	   \includegraphics[width=0.5\columnwidth]{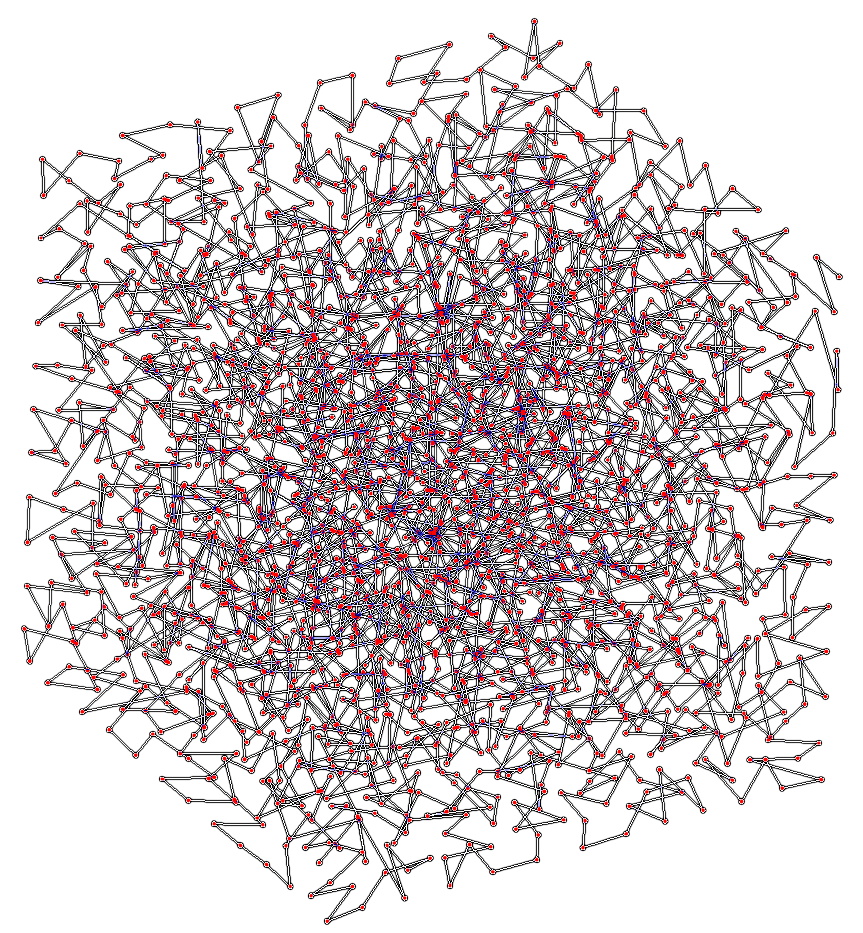}
	   \includegraphics[width=0.5\columnwidth]{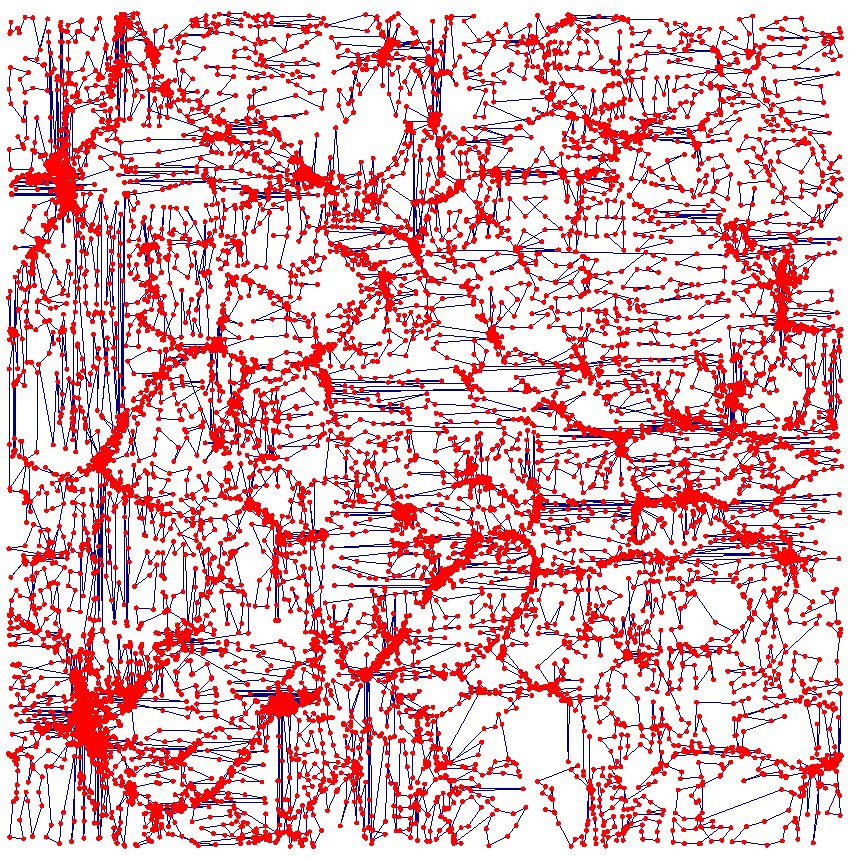}
    }
    \centerline{
       C \hspace{0.5\columnwidth} D 
    }
    \caption{A: the Hilbert curve. B: a 2D Hilbert-ordered pointset. 
       C: a 3D Hilbert-ordered pointset. D: adaptive Hibert ordering using the median,
       resulting in a balanced Hilbert curve, adapted to an heterogeneous pointset.}
    \label{fig:hilbert}
\end{figure}

The only thing we need to explain now is how to compute a Laguerre diagram.

\subsection{Geometrical aspects}

To compute the Laguerre diagram, we use the classical algorithm developed simultaneously by Bowyer and Watson \citep{DBLP:journals/cj/Bowyer81,journals/cj/Watson81} (initially for Voronoi diagrams). We do not completely detail this algorithm, but give the general idea below. Then we mention some specificities of our implementation.

\subsubsection*{Bowyer-Watson and the dual triangulation}

While it would be possible to directly represent the polygons / polyhedra of the Laguerre diagram, it would be costly, because each polygon/polyhedron can have a different number of vertices. The Bowyer-Watson algorithm uses the dual triangulation instead, displayed in gray in Figure \ref{fig:triangulation}: each Laguerre vertex is shared by three Laguerre cells. Thus, what is represented in the computer is the set of $(i,j,k)$ indices triplets such that the Laguerre cells $V^\psi_i$, $V^\psi_j$
and $V^\psi_k$ have a common vertex. This forms a triangulation of the pointset $(\bx_i)_{i=1}^N$, known as the \emph{regular triangulation} \citep{DBLP:journals/siamcomp/Aurenhammer87}. In 3D, each Laguerre vertex is shared by four Laguerre cells, and the dual structure is made of $(i,j,k,l)$ tetrahedra (instead of triangles in 2D). \\

This triangulation is constructed by inserting the points $\bx_i$ one by one. Each time a point $\bx_i$ is inserted, the triangles/tetrahedra that correspond to the Laguerre vertices that fall inside the cell $V^\psi_i$ of $\bx_i$ are discarded, and the triangles that correspond to the vertices of $V^\psi_i$ are created. The Boywer-Watson algorithm uses the fact that the set of triangles to be discarded is connected, and comprises the triangle that contains $\bx_i$. This remark makes it possible to speed-up the algorithm: starting from the triangle that contains $\bx_i$, found by navigating the triangulation, a greedy algorithm traverses the set of triangles to be discarded. This dramatically reduces the number of triangles to be tested.

\subsubsection*{Spatial sorting}

At this point, the execution time is dominated by finding the triangle/tetrahedron that contains each point $\bx_i$. In 3D, starting from a random tetrahedron, the algorithm needs to traverse an everage of $\sqrt[3]{N}$ tetrahedra to find the one that contains $\bx_i$. \\

The algorithm is made significantly faster by sorting the vertices spatially \citep{Amenta:2003:ICC:777792.777824,DBLP:conf/imr/AlauzetL09}, along the Hilbert curve (see Figure \ref{fig:hilbert}-A). Sorting the vertices this way ensures that two points near to each other in 3D are mapped to close indices. Hilbert sorting is classical in high performance large scale cosmological simulation, for instance, it is a key component of the code used in the DEUS project \citep{DBLP:journals/ijhpca/ReverdyABRCBRDR15}. Figure \ref{fig:hilbert}-B,C shows what the computed order looks like for a homogeneous point distribution. In our case, the distribution of points can be highly heterogeneous, with a large number of points clustered in some zones. As suggested in \cite{WEB:SpatialSorting}, to make the ordering well adapted to the point distribution, we use the median of the points coordinates to hierarchically subdivide the domain, see Figure \ref{fig:hilbert}-D for a 2D example. \\

Each time a new point is inserted, the tetrahedron that contains it is searched by navigating the triangulation starting from a tetrahedron incident to the previously inserted point. Since points with consecutive indices are near to each other in 3D, this considerably reduces the number of traversed tetrahedra (from $\sqrt[3]{N}$ to typically 10-20).  \\

Spatial sorting not only accelerates the computation of the Laguerre diagram, but also it speeds-up the iterative conjugate gradient solver: since it maps neighboring points to as-contiguous-as-possible locations in memory, it significantly improves cache locality. Without spatial sorting, on the GPU, we obtain 70 GFlops without it and 90 GFlops with it. On the CPU, we obtain 1.5 GFlops without it and 2 GFlops with it.

\subsubsection*{Numerical precision and geometric predicates}

To determine which tetrahedron needs to be created or discarded, the algorithm needs to take combinatorial decisions based on the relative locations of some geometric elements. So the algorithm depends on a limited
number of functions, called \emph{geometric predicates}. A predicate is a function that takes as arguments a set of points $\bx_i, \bx_j, \bx_k \ldots$ and $\psi_i, \psi_j, \psi_k \ldots$ coefficients, and that returns 
a discrete value $-1, 0$ or $+1$. For computing a Laguerre diagram, we need two geometric predicates: 
\begin{itemize}
\item  ${\tt orient}$, that indicates whether the three vectors 
       $(\bx_j - \bx_i, \bx_k - \bx_i, \bx_l - \bx_i)$ forms a direct (+1), degenerate (0) or indirect (-1) basis. It is used to navigate the triangulation and find the tetrahedron that contains $\bx_i$: 
$$
   {\tt orient(\bx_i, \bx_j, \bx_k, \bx_l)} = \mbox{sign } \begin{vmatrix} x_j - x_i & y_j - y_i & z_j - z_i \\ x_k - x_i & y_k - y_i & z_k - z_i  \\ x_l - x_i & y_l - y_i & z_l - z_i  \end{vmatrix};
$$
       
\item ${\tt conflict}$, that indicates whether the Laguerre vertex that corresponds to the tetrahedron $(j,k,l,m)$ falls inside the Laguerre
   cell $V^\psi_i$ of $\bx_i$ (+1), on its boundary (0) or outside (-1). It is used to determine which tetrahedra need to be discarded when inserting $\bx_i$ in the diagram:
\ba
\nonumber
   {\tt conflict}(\bx_i, \bx_j, \bx_k, \bx_l, \bx_m, \psi_i, \psi_j, \psi_k, \psi_l, \psi_m) = \\
\nonumber
   \mbox{sign }
   \begin{vmatrix} x_j - x_i & y_j - y_i & z_j - z_i & h_j - h_i \\ x_k - x_i & y_k - y_i & z_k - z_i & h_k - h_i \\ 
                   x_l - x_i & y_l - y_i & z_l - z_i & h_l - h_i \\ x_m - x_i & y_m - y_i & z_m - z_i & h_m - h_i \end{vmatrix}, \\
    \nonumber
    \mbox{where  } h_i = x_i^2 + y_i^2 + z_i^2 - \psi_i \mbox{ (resp. } j,k,l,m \mbox{)}.
\ea
\end{itemize}

The two predicates ${\tt orient}$ and ${\tt conflict}$ correspond to the sign of polynomials of the points coordinates and coefficients of $\psi$. It is of crucial importance that these signs are coherent: for instance, if at one moment the algorithm considers that point $\bx_i$ is strictly above point $\bx_j$, the algorithm should not consider later that $\bx_j$ is strictly above $\bx_i$, else it will create a triangulation that is not coherent. Since floating point numbers have a limited precision, avoiding this type of inconsistencies requires special care. It is especially true in our case, since we are computing a large number of Laguerre diagrams (typically tenths) with a huge number of points (typically tenths millions). In the 2000's, it was a major obstacle to the early development of cosmological codes with the semi-discrete setting in 3D.  To ensure that the combinatorial decisions taken by the algorithm are coherent, we developed the PCK (Predicate Construction Kit) programming language \citep{DBLP:journals/cad/Levy16}, that transforms the formula of a predicate into a function that evaluates the exact sign. We used it to implement {\tt orient}, {\tt conflict} and other specialized predicates \citep{DBLP:conf/gmp/YanWLL10} involved in the periodic boundary condition (next paragraph). Internally, we use exact expansion-based arithmetics \citep{DBLP:journals/dcg/Shewchuk97}, that represent each number by an array of double-precision floating point numbers. To speed-up computations, we also use arithmetic filters \citep{meyer:inria-00344297}, that quickly determines the signs in the easy cases and avoid costly expansion-arithmetics computations in most cases. Finally, we use symbolic perturbation \citep{Edelsbrunner90simulationof} to ensure that the decisions remain coherent even in degenerate configurations. 

\begin{figure}
    \centerline{
	   \includegraphics[width=0.5\columnwidth]{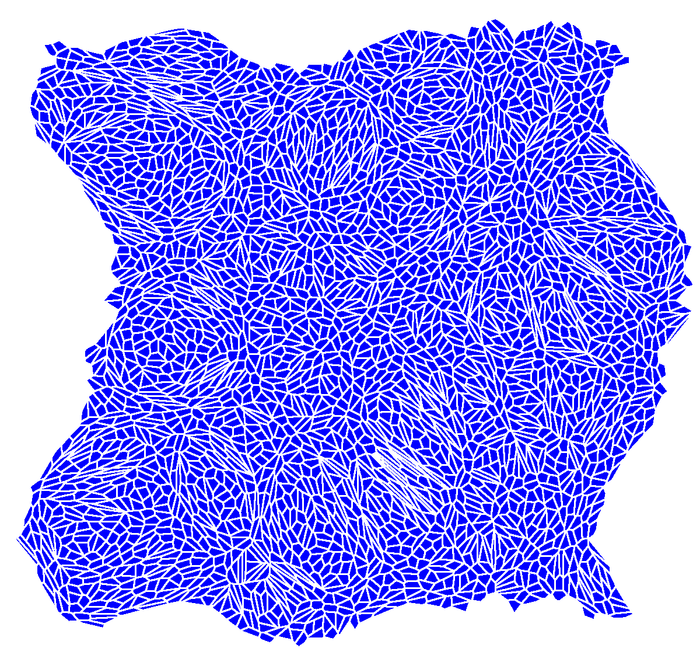}
	   \includegraphics[width=0.5\columnwidth]{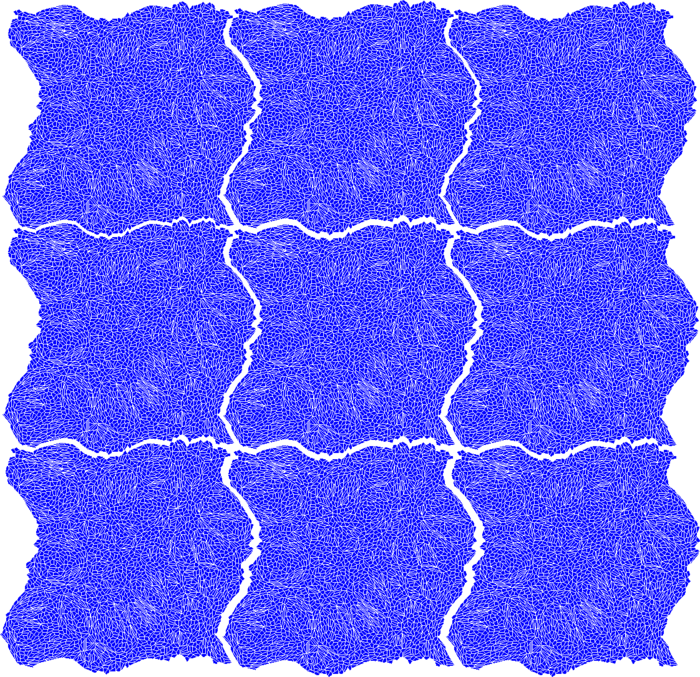}
    }
    \centerline{
       A \hspace{0.5\columnwidth} B 
    }
    \centerline{
 	   \includegraphics[width=0.75\columnwidth]{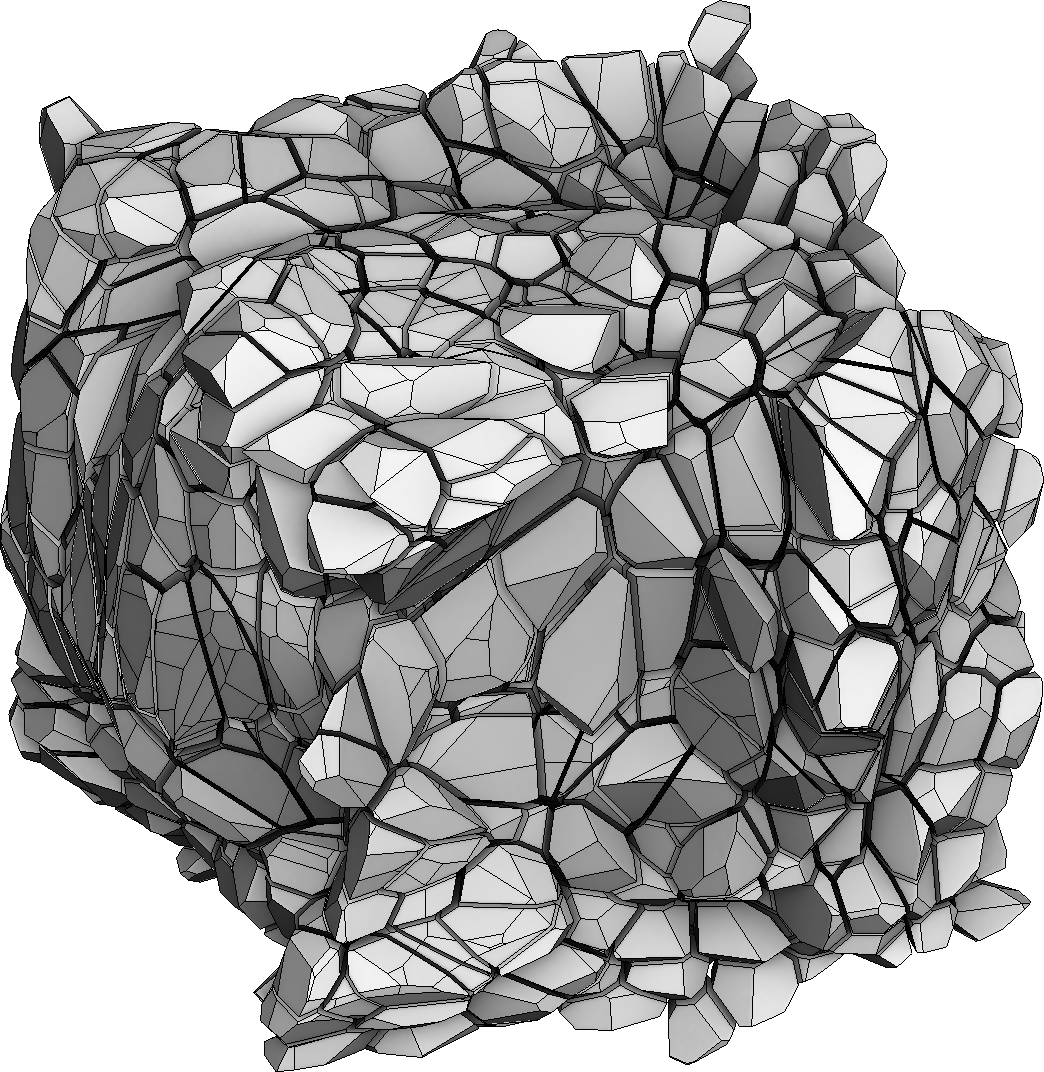}
    }
    \centerline{ C }
    \caption{Laguerre diagrams with periodic boundary conditions in 2D and in 3D.}
    \label{fig:periodic}
\end{figure}

\subsubsection*{Periodic boundary conditions}

Remember that our computational domain $V$ is the unit cube $[0,1]^3$ with periodic boundary conditions. With earlier discrete-discrete methods, like the "auctions" algorithm, it is easy to take this into account: each time the squared distance $| \bx_i - \bx_j |^2$ needs to be computed, it is replaced with:
$$
\mbox{min}_{k=1}^{27} | \bx_i - T_k(\bx_j) |^2,
$$
where $T_k$ denotes one of the 27 possible translations obtained using -1,0,1 coordinates. \\

In our case, the situation is more complicated, because we need to compute the (continuous) Laguerre diagram. A trivial solution consists in copying the points 27 times (and keep only the center part with zero translation). Clearly this would dramatically increase the computation time. What we do instead is first computing the Laguerre diagram of the points, then determining which cell intersects the boundary of the $[0,1]^3$ cube, and copy these points with the right translations, depending on which face, edge or vertex of the cube was intersected \citep{DBLP:conf/isvd/YanWLA11}. Note that the (back-translated) neighbors of these translated points need to be inserted as well. This typically concerns 5 to 10\% of the points (1.1 times, to be compared with 27 times). Figure \ref{fig:periodic}-A shows an example of a 2D periodic Laguerre diagram. The diagram paves the 2D space (Figure \ref{fig:periodic}-B). Figure \ref{fig:periodic} shows a 3D example (that paves the 3D space). 

For the sake of completeness, we also mention the alternative approach in \citep{DBLP:conf/esa/CaroliT09} for computing periodic Laguerre diagrams, that is more elegant theoretically. It iteratively inserts the vertices, starting with 27 copies, then it switches to a periodic triangulation where each vertex and tetrahedron is only represented once, with combinatorics that represent the periodic space, as soon as a criterion on the points location and $\psi_i$'s is respected. However, while their approach works well in practice for Voronoi diagrams (with $\psi$ = 0), their criterion is never met in our case, resulting in 27 copies, because the $\psi_i$ coefficients vary too much. Moreover, even with a single copy, the memory consumption of the data structures they use (CGAL library, pointer-based) makes it not practical for large-scale cosmological simulations. 

\subsubsection*{Multicore}

The Bowyer-Watson algorithm is not well suited to multicore parallelisation, because it inserts the points one by one in a way that globally updates the diagram under construction. However, still using spatial sorting, it is possible to split the pointset into batches that are geometrically well separated and unlikely to interact. Each batch is assigned to a different thread. Then all the threads insert their batches of points into the diagram simultaneously. We use light weight synchronization primitives (spinlocks) to detect whenever two threads try to modify the same tetrahedron. Such conflicts are resolved by rolling-back the involved modifications of the diagram, and redoing them in sequential mode. In our computations, for a 80 million points diagram, typically a few tenth of conflicts are encountered (performance penalty is negligible).

\section{Lagrangian to Eulerian conversion}

\label{sec:LagrangeEuler}

\begin{figure}
    \centerline{
	   \includegraphics[width=\columnwidth]{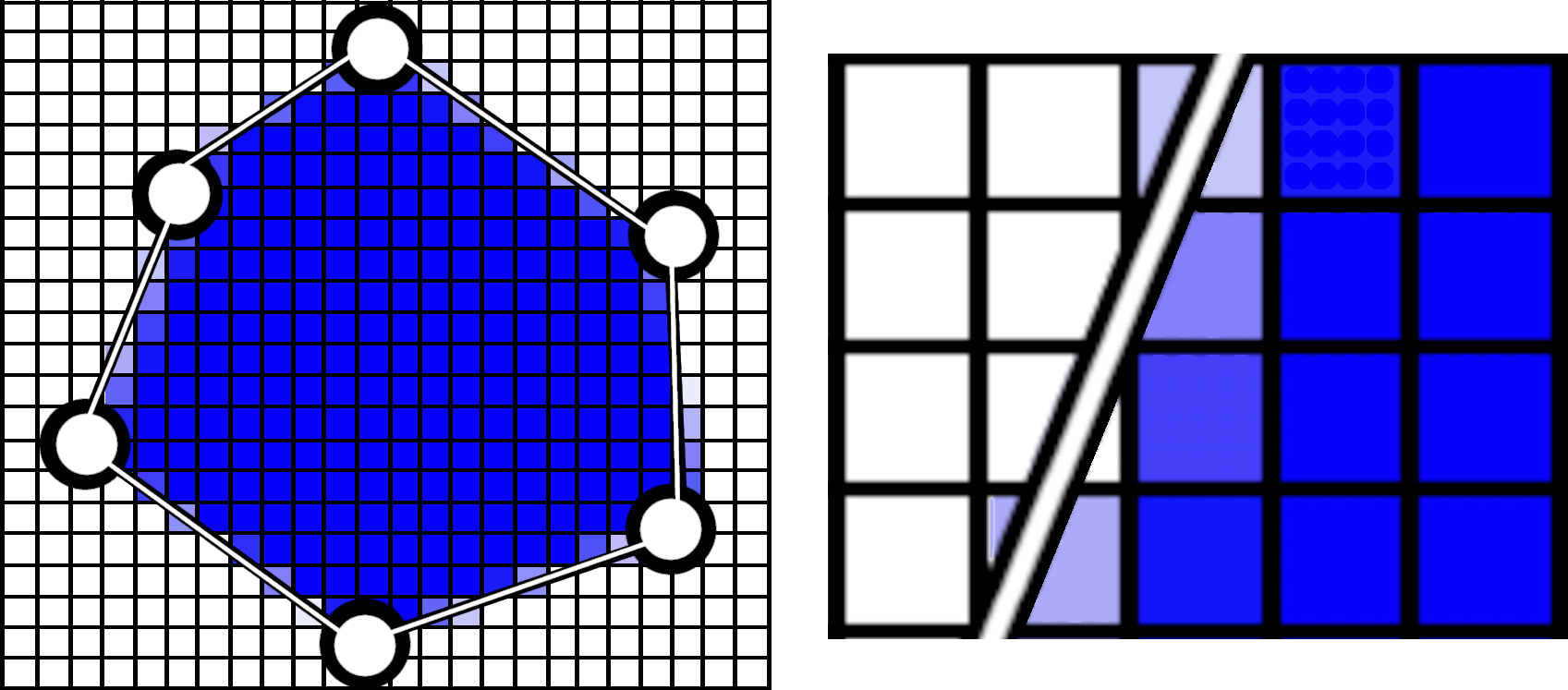}
    }
    \vspace{5mm}
    \centerline{
	   \includegraphics[width=\columnwidth]{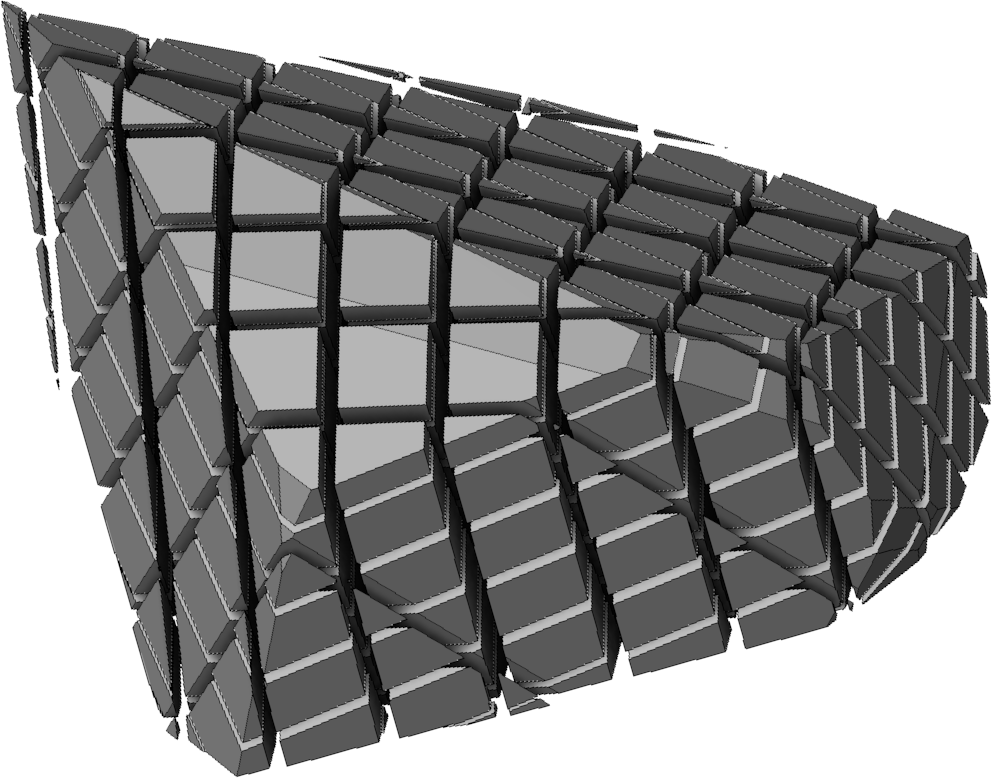}
    }
    \caption{
       Converting from Lagrangian (Laguerre cell) to Eulerian (density grid) by measuring the area of the intersection between the Laguerre cell and the grid cells, in 2D (top) and in 3D (bottom).
    }
    \label{fig:laguerre_raster}
\end{figure}

The Lagrangian-to-Eulerian conversion means 'painting' the Laguerre cells into the regular grid (see Figure \ref{fig:laguerre_raster}), which can be done at any redshift $z_s$ by moving the vertices of the Laguerre cell using Eq. (\ref{eq:zeldovichapprox2}) page \pageref{eq:zeldovichapprox2}. All the cells of the Laguerre diagram are subsequently 'painted' onto the grid in parallel, by determining the regular grid cells contained by each Laguerre cell. Along the boundary of the shrunken Laguerre cell, we compute the intersection volume between the Laguerre cell and the grid cells. Since both objects (Laguerre cell and grid cell) are convex, their intersection can be easily computed using a dual representation, see \cite{DBLP:journals/siamcomp/Aurenhammer87} and references herein for more details.

This Lagrangian-to-Eulerian conversion typically takes 30 minutes for converting a 16 million cells Laguerre diagram into a $512^3$ Eulerian density grid.

\section{Details of power spectrum preparation}
\label{app:psdetails}

\subsection{Shot noise}
\label{app:psdetails_shotnoise}

Shot noise generally refers to Poisson noise of a discretised continuous density field and in most practical examples scales inversely proportional to the number of discretising sources $N$.  The power spectrum corresponding to the particle distribution in question is thus increased by the $k$-independent noise contribution as $P(k,N) = P(k)+P_{sn}(N)$.  The shot noise scales as
\begin{equation}
    P_{sn}(N) = A\cdot N^{-1},
    \label{eq:shotnoise}
\end{equation}
where $A=l^3$ is the volume of the box.  However, the power spectra shown in Figure~\ref{fig:powerspectrum} were not computed from a discrete set of point particles.  Instead, the continuous density field was effectively discretised by extended objects, the shrunken Laguerre cells, e.g.~Figure~\ref{fig:periodic}.  It is non-trivial to predict the resulting shot noise for these structures.  Empirically, however, Equation~\ref{eq:shotnoise} seems to describe the shot noise contribution of this arrangement rather well, simply after promoting $A$ to be a free parameter.  
The following paragraph describes how the shot noise is computed and finally subtracted from the corresponding power spectra in Figure~\ref{fig:powerspectrum}.

\begin{figure*}
    \centering
    \includegraphics[width=\columnwidth]{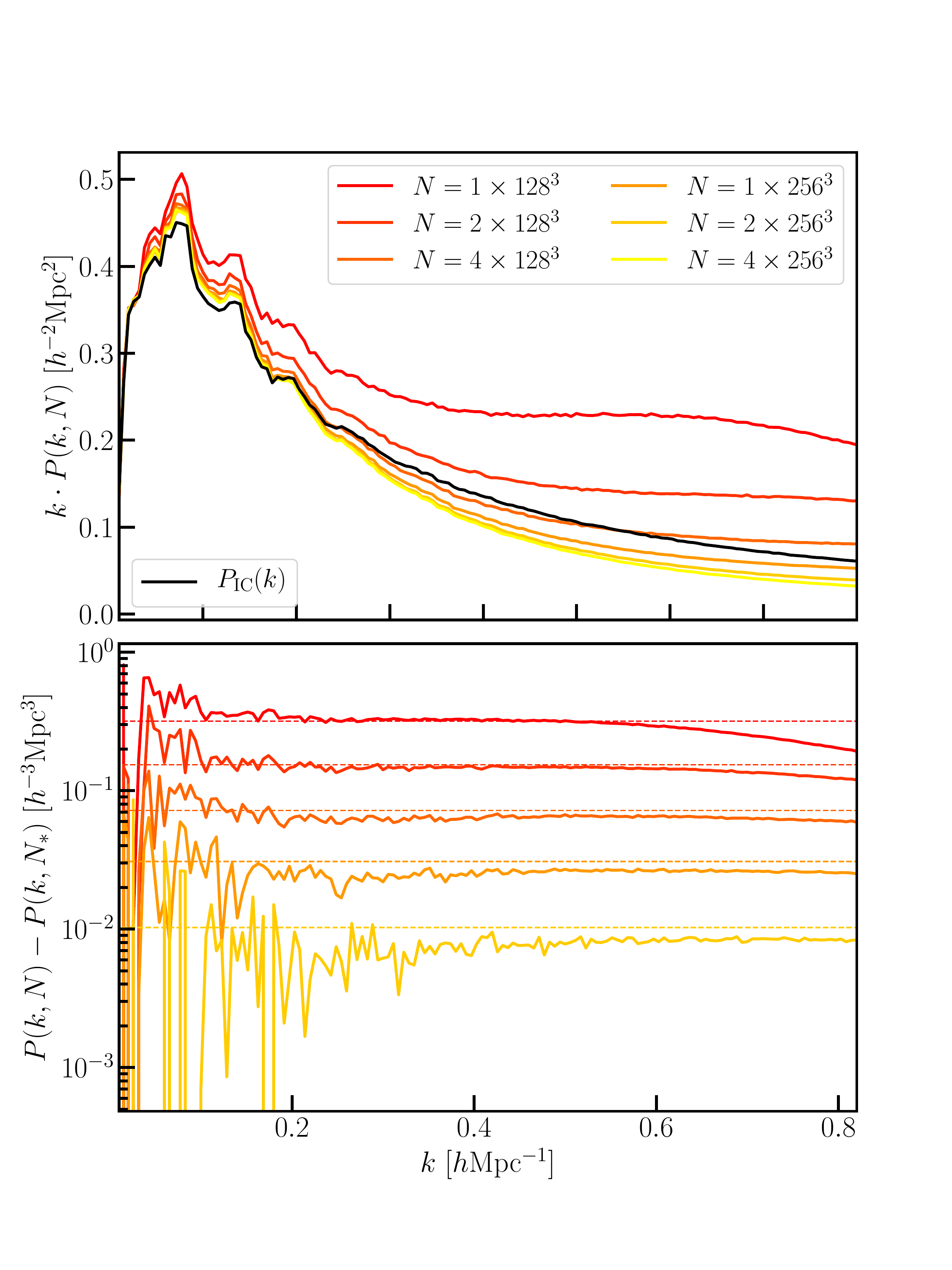}
    \includegraphics[width=\columnwidth]{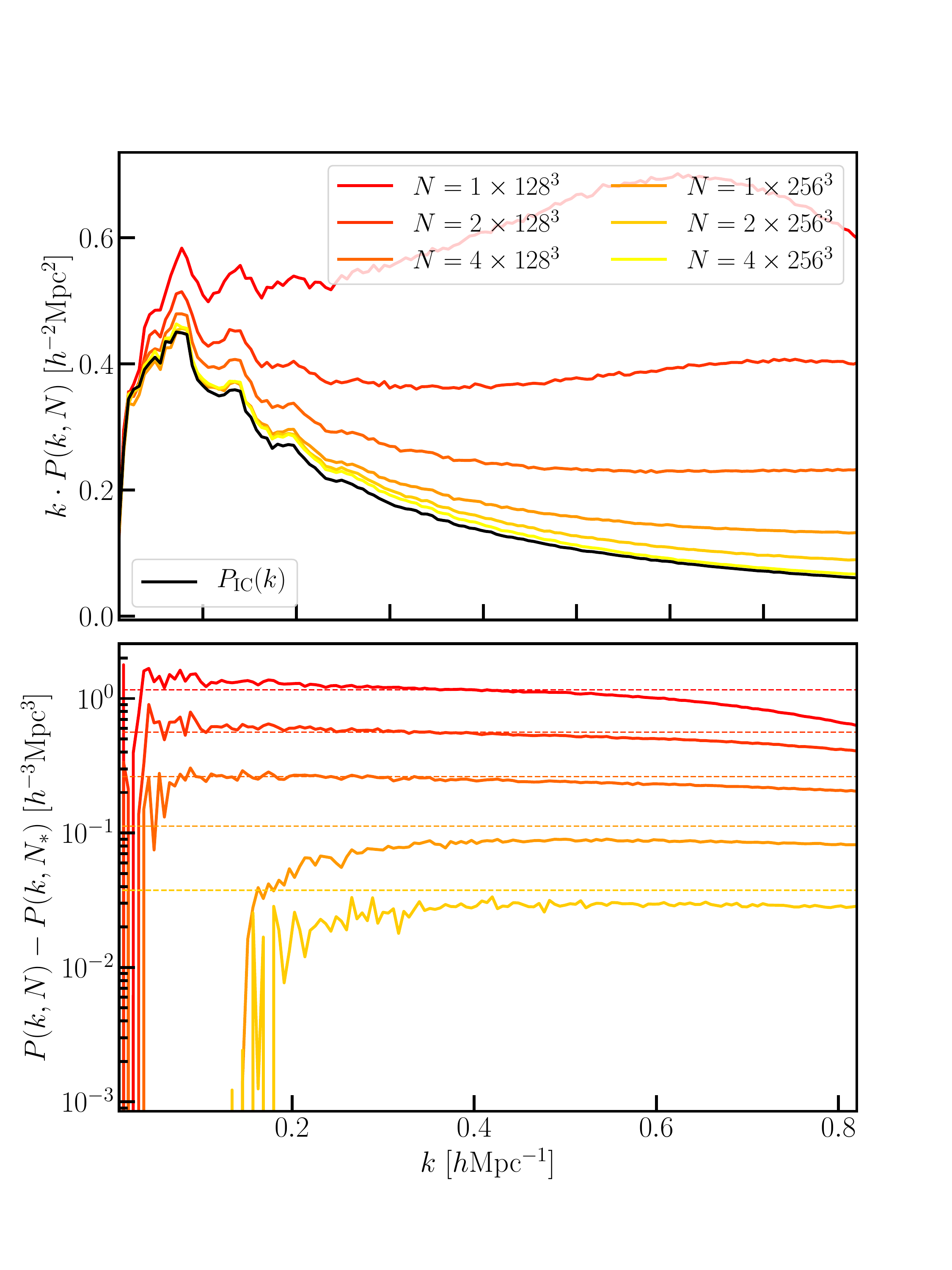}
    \caption{\textit{Upper panels:} Power spectra of reconstructions using particle samples of different size $N$, highlighting the $N$ dependent shot noise.  Reconstructions were done with an input redshift of $z=0.3$ (\textit{left}) and $z=1.5$ (\textit{right}).  \textit{Lower panels:} Differences of each of the power spectra with that of the largest sample size, $N=4\times 256^3=67,108,864$.}
    \label{fig:shotnoise}
\end{figure*}

By definition of what should here be referred to as shot noise, the scaling with particle number $N$ allows an estimation of its amplitude $A$ from comparing reconstructed power spectra with one another, without the knowledge of the desired, ``true'' power spectrum that provided the initial condition (IC).  This is especially promising for future reconstructions using real data, where the true solution is obviously unknown.  We perform reconstructions of the initial density at $z=49$ as in the main body, using particle samples of increasing size.  Their uncorrected power spectra are shown in the upper panels of Figure~\ref{fig:shotnoise} along with the initial condition power spectrum, clearly demonstrating the presence of shot noise.  Without reference to the IC power spectrum, the shot noise amplitude of Equation~\ref{eq:shotnoise} is fit by comparing power spectra of different particle number with that of one selected particle number $N_*$:
\begin{equation}
    P(k,N)-P(k,N_*) = A\cdot\frac{N_*-N}{NN_*}
    \label{eq:shotnoise_difference}
\end{equation}
These quantities are shown in the bottom panels of Figure~\ref{fig:shotnoise} with $N_*=2\times256^3$, the largest particle sample used here, and clearly show the corresponding $N$-dependent contributions to the reconstructed power spectra.  It should be stated here, that any discrepancies that remain between the reconstructed power spectra after proper shot noise subtraction and the IC power spectrum are therewith shown \textit{not} to be dependent on the size of the particle sample, and must be accounted for by other means.  In each of the same panels, indicated by dashed lines, the relation~\ref{eq:shotnoise_difference} was fitted collectively to all curves within the range $0.25\,h\textrm{Mpc}^{-1}<k<0.5\,h\textrm{Mpc}^{-1}$.  For the reconstructions from $z=0.3$ and $1.5$ (left and right panels) we find:

$$
\begin{array}{c|cc}
                      &  z_i=0.3 & z_i=1.5 \\
\hline
A\;[h^{-3}\rm{Mpc}^3] &   660,078.73 & 2,327,572.13      
\end{array}
$$

Finally, this allows for calculating the shot noise for a reconstruction from a given particle number via Equation~\ref{eq:shotnoise}.  For the particle sample of size $N=256^3$, as employed in the main body of this paper, we find

$$
\begin{array}{c|cc}
   & z_i=0.3  & z_i=1.5 \\
\hline   
P_{sn}(N=256^3) & 0.0393   & 0.1387   
\end{array}
$$

Note that, both values are smaller than the shot noise expected for a discrete particle sample, even after accounting for a scaling by the linear growth factors, $D^2(z=49)/D^2(z_i)\cdot l^3/N$.

\subsection{Subsample variance}
\label{app:psdetails_subsamplevariance}

\begin{figure*}
    \centering
    \includegraphics[width=\columnwidth]{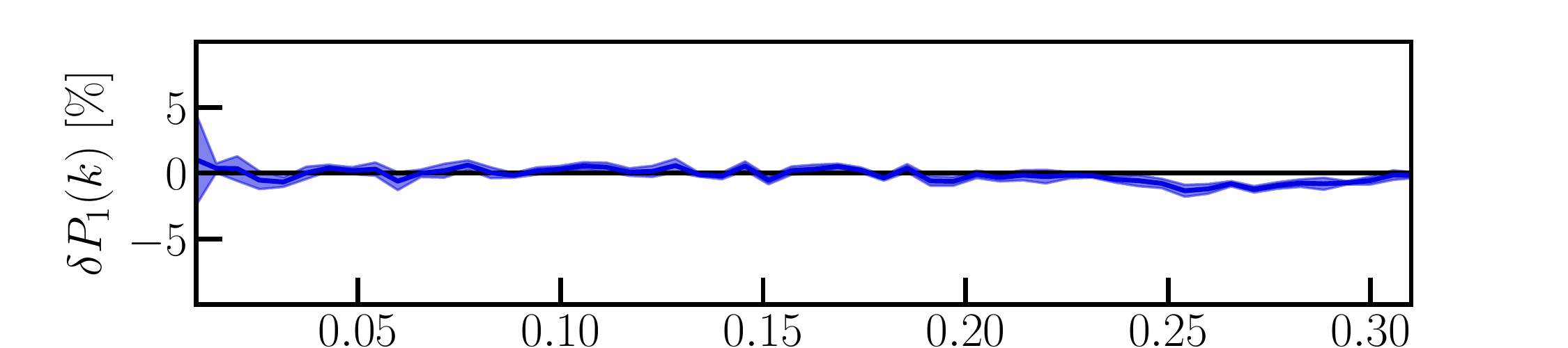}
    \includegraphics[width=\columnwidth]{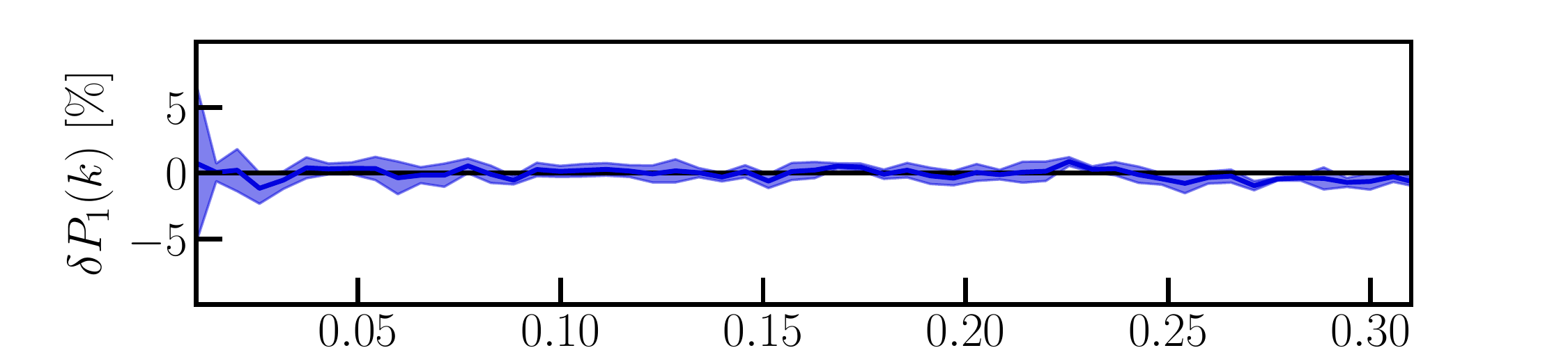}
    \caption{Illustration of subsample variance given our choice of $N=256^3$ particles.  Shown are mean and standard deviation of the relative difference $\delta P_1(k)$, Eq.~\ref{eq:powerspectrumdeviation}, for simulation $i=1$ averaged over all five subsamples $j$, at input redshifts of $z=0.3$ (\textit{left}) and $z=1.5$ (\textit{right}).}
    \label{fig:subsamplevar}
\end{figure*}

Subsample variance decreases with increasing particle number. Sampling as many as $256^3$ particles therefore suppresses the influence of these terms considerably.  Nevertheless, we show in Fig.~\ref{fig:subsamplevar} the relative difference of the power spectra of each of one simulations' five subsamples and their average. The variance on average makes up about 75\% of the total variance shown in Figs.~\ref{fig:powerspectrum}--\ref{fig:powerspectrum_fitted} in the range shown.

\bsp	
\label{lastpage}
\end{document}